\documentclass[11pt,a4paper,english,reqno]{amsart}

\title[Minimum bounding polytropes for MLBNs]
      {Minimum bounding polytropes for estimation of max-linear Bayesian networks}

\author{Kamillo Ferry}
\address{Kamillo Ferry, Technical University Berlin, Germany}
\email{ferry@math.tu-berlin.de}

\date{}


\usepackage[fixamsmath,disallowspaces]{mathtools}
\usepackage{amsmath}
\usepackage{amssymb}
\usepackage{amsthm}
\usepackage{stmaryrd}
\usepackage{faktor}

\usepackage{hyperref}

\usepackage{booktabs}
\usepackage{graphicx}
\usepackage{standalone}
\usepackage{subcaption}
\usepackage[mathscr]{euscript}

\usepackage{setspace}
\usepackage[headings]{fullpage}

\usepackage[dvipsnames]{xcolor}

\usepackage{algorithmicx}
\usepackage{algorithm}
\usepackage[noend]{algpseudocode}

\usepackage{tikz}
\usetikzlibrary{decorations.pathreplacing}
\usetikzlibrary{positioning,fit}
\usetikzlibrary{graphs,graphs.standard}
\usetikzlibrary{cd}
\usetikzlibrary{shapes}
\usetikzlibrary{arrows,arrows.meta}
\usetikzlibrary{chains}
\usetikzlibrary{fit}
\usetikzlibrary{calc}
\usetikzlibrary{patterns,patterns.meta}

\usepackage{pgfplots}
\pgfplotsset{compat=1.18}

\usepackage{subcaption}

\usepackage{amsmath}
\usepackage{amsthm}
\usepackage{amssymb}
\usepackage{mathtools}

\usepackage{blkarray}

\usepackage[
  backend=biber,
  style=numeric,
  maxbibnames=99,maxcitenames=99,
  backref,backrefstyle=none
]{biblatex}
\newcommand{\citeauthorandcite}[2][]{\citeauthor{#2}~\cite[#1]{#2}}

\usepackage[nameinlink]{cleveref}

\theoremstyle{plain}
\newtheorem{theorem}{Theorem}
\numberwithin{theorem}{section}
\newtheorem{proposition}[theorem]{Proposition}
\newtheorem{lemma}[theorem]{Lemma}

\theoremstyle{definition}
\newtheorem{definition}[theorem]{Definition}
\newtheorem{example}[theorem]{Example}

\theoremstyle{remark}
\newtheorem{remark}[theorem]{Remark}

\newtheorem{question}{Question}

\usepackage[nameinlink]{cleveref}

\newcommand{\fundamentalpolytope}[1][\thedag]{\ensuremath{P_{#1}}}

\newcommand{\numberClassFont}[1]{\protect\ensuremath{\mathbb{#1}}}     
\newcommand{\RR}{\numberClassFont{R}}

\newcommand{\NN}{\numberClassFont{N}}

\newcommand{\TT}{\numberClassFont{T}}
\newcommand{\TTAA}{\TT\numberClassFont{A}}
\newcommand{\TTPP}{\TT\numberClassFont{P}}

\newcommand{\thedag}{\ensuremath{G}}
\newcommand{\subdag}{\ensuremath{H}}

\def\wdp(#1){\ensuremath{\mathrm{Q}(#1)}}
\def\WDP(#1)i{\wdp(#1)}
\def\tightspan(#1){\wdp(#1)}

\newcommand{\subdivision}{\ensuremath{\Sigma}}

\newcommand{\estimate}{\widehat\thedag}
\newcommand{\estimateEdges}{\widehat{E}}
\newcommand{\estimateC}{\widehat{C}}

\newcommand{\bigo}{\mathcal{O}}

\newcommand{\conv}{\ensuremath{\mathrm{conv}}}
\newcommand{\tconv}{\ensuremath{\mathrm{tconv}}}

\newcommand{\SymGr}{\ensuremath{\mathbb{S}}}

\colorlet{maincolorDark}{gray}

\algnewcommand\algorithmicOutput{\textbf{Output:}}
\algnewcommand\Output{\item[\algorithmicOutput]}
\algnewcommand\algorithmicInput{\textbf{Input:}}
\algnewcommand\Input{\item[\algorithmicInput]}

\algnewcommand\algorithmicswitch{\textbf{switch}}
\algnewcommand\algorithmiccase{\textbf{case}}

\algnewcommand\algorithmicassert{\texttt{assert}}
\algnewcommand\Assert[1]{\State \algorithmicassert(#1)}%
\algnewcommand\algorithmicParameter{\textbf{Parameter:}}
\algnewcommand\Parameter{\item[\algorithmicParameter]}

\algdef{SE}[SWITCH]{Switch}{EndSwitch}[1]{\algorithmicswitch\ #1\ \algorithmicdo}{\algorithmicend\ \algorithmicswitch}%
\algdef{SE}[CASE]{Case}{EndCase}[1]{\algorithmiccase\ #1:}{\algorithmicend\ \algorithmiccase}%
\algtext*{EndSwitch}%
\algtext*{EndCase}%

\begin{document}
\onehalfspace

\begin{abstract}
    Max-linear Bayesian networks are recursive max-linear structural equation models
    represented by an edge weighted directed acyclic graph (DAG). 
    The identifiability and estimation of max-linear Bayesian networks is an intricate
    issue as \citeauthor{GKL:2021} have shown. As such, a max-linear Bayesian network is generally 
    unidentifiable and standard likelihood theory cannot be applied.

    We can associate tropical polyhedra to max-linear Bayesian networks. Using this, we investigate the minimum-ratio estimator
    proposed by \citeauthor{GKL:2021} and give insight on the structure of minimal best-case samples for parameter recovery
    which we describe in terms of set covers of certain triangulations.
    
    We also combine previous work on estimating max-linear models from \citeauthor{BKT:2024}
    to apply our geometric approach to the structural inference of max-linear models.
    This is tested extensively on simulated data and on real world data set, the NHANES report for 2015--2016
    and the upper Danube network data.
\end{abstract}

\maketitle

\section{Introduction}
Graphical models are a powerful tool for modeling the dependencies of multivariate random vectors, popular for their 
graph representation and subsequent intuitive representation of causal relations. In \citeyear{GK:2018}, \citeauthorandcite{GK:2018}
introduced \emph{max-linear Bayesian networks} (MLBN) to describe the causal relations of variables where large values 
propagate through the network. They have already been successfully applied to modeling how flooding events propagate through
river networks \cite{ADE:2015,BKT:2024} and also to study tail dependencies in data from the European stock market 
\cite{EKS:2018}.

Max-linear Bayesian networks are random vectors \(X = (X_1,\dots,X_d)\) specified by a DAG \(\thedag = ([d],E)\) 
with weight matrix \(C\in\RR_{\geq0}^{d\times d}\) and independent random variables \(Z_1,\dots,Z_d\) called \emph{innovations}.
An MLBN satisfies the following recursive \emph{structural equations}
\begin{equation}\label{eq:recursive-structural-eqn}
    X_i = \bigvee_{j=1}^n c_{ij}X_j \vee Z_i	
\end{equation} over the max-times semiring where \(\vee\) denotes taking the maximum of the operands.

Of great importance for the application of MLBNs to real-world problems is the question of inferring the structure and
parameters from sample data. Previous work by \citeauthorandcite{GKL:2021} has shown that
an MLBN to be generally unidentifiable while characterizing the class of edge weights that could be recovered from
a given sample and giving an estimator for a distinguished element of this class.
Subsequently, the focus has then been on estimation of the edge weights for an MLBN under various assumptions, e.\,g.\ 
assuming regularly varying distribution tails for the innovations \cite{KK:2021}, or estimation using noisy measurements
through scoring \cite{BK:2021,BKT:2024} or Gaussian mixtures \cite{AFY:2025}.

Being defined over the max-times semiring, there is a rich connection between MLBNs and tropical geometry which has been outlined 
by \citeauthorandcite{Tran:2022} and that has been recently investigated by the author \cite{AF:2024}. 
Using this connection, we describe the geometrical meaning of the 
minimum-ratio estimator proposed by \citeauthorandcite{GKL:2021} and describe combinatorial properties of it. 
This gives rise to the question of set covers of certain subdivisions of subpolytopes of \(A_d\) root polytopes
which is a combinatorial problem that can be studied in its own right.

We also combine this estimator with techniques from \citeauthorandcite{BK:2021} and \citeauthorandcite{BKT:2024} 
to approach the problem of structural inference for MLBNs.
In addition, we conduct several computational experiments for which me make the data available under 
\url{https://zenodo.org/records/17054748}.

\subsection*{Outline}
\Cref{sec:preliminaries} contains a summary of tropical convexity and polytropes, their combinatorial properties and
connection to regular subdivisions of certain point configurations and basic properties of max-linear Bayesian networks.
In \Cref{sec:minimum-bounding-polytrope} we consider the minimum-ratio estimator from \citeauthorandcite{GKL:2021}
after applying a logarithmic transformation and describe the tropical geometric interpretation as minimum bounding polytrope
of a given sample.
In \Cref{sec:param-recovery} we describe the properties of minimal best-case samples for which the minimum bounding polytrope
can correctly identify the weight matrix when the underlying DAG is known and make a connection to minimum set covers 
of certain triangulations.
In \Cref{sec:structural-learning} we then move towards the setting where the underlying DAG is unknown and
combine the minimum bounding polytrope with scoring and thresholding methods to give an algorithm for structural inference.
In \Cref{sec:experiments} we apply our method to simulated and real-world data and discuss its performance.
We conclude with an outlook on the question of minimal set covers for the central triangulations of root polytopes.

\section{Preliminaries}\label{sec:preliminaries}
\subsection{Tropical convexity}
Denote by \(\TT \coloneqq (\RR\cup\{\infty\}, \oplus = \min, \odot=+)\) the \emph{tropical (min-plus) semiring}
where \(\oplus = \min\) and \(\odot=+\). We define the \emph{tropical affine space} \(\TTAA^{d-1}\) as the 
set of points \(x\in\RR^d\) with the equivalence relation given by (tropically) scaling, that is 
\[(x_1,\dots,x_n)\sim\lambda\odot(x_1,\dots,x_d)=(x_1+\lambda,\dots,x_d+\lambda)\] for all \(\lambda\in\RR\).
Then, the \emph{tropical projective space} \(\TTPP^{d-1}\) is defined as \[
    \TTPP^{d-1} \coloneq \faktor{\left(\TT^{d}\setminus\{(\infty,\dots,\infty)\}\right)}{\mathbb{R}\mathbf{1}}
\] where \(\mathbf{1} = (1,\dots,1)\). 
This is the compactification of \(\TTAA^{d-1}\) in the sense that the pair
\((\TTPP^{d-1},\TTAA^{d-1})\) is homeomorphic to the standard \(d\)-simplex and its interior.

Fix a finite set of \(n\) points \(C\subset\TTPP^{d-1}\). By abuse of notation, we write \(C\) for this set 
and the \(d\times n\)-matrix over \(\TT\) whose columns are the points \(c^{(i)}\in C\). 
Additionally, we require that no row sum of \(C\) is equal to \(\infty\).

The \emph{tropical polytope} \(\tconv(C)\) spanned by \(C\) is then defined as the set of \(\RR\)-linear 
combinations of the points in \(C\), or equivalently, as the image of the map \(\TTAA^{n-1}\to\TTAA^{d-1}\)
given by left multiplication with \(C\).
Tropical polytopes are \emph{tropically convex} in the sense that for any two points \(p,q\in P\)
in a tropical polytope \(P\subseteq\TTAA^{d-1}\), the \emph{tropical line segment} \(\tconv(\{p,q\})\)
is entirely contained inside \(P\).

The tropical projective torus \(\TTAA^{d-1}\) may be identified with Euclidean space \(\RR^{d-1}\)
normalizing a coordinate to \(0\). Since the equivalence relation on \(\TTAA^{d-1}\) allows us to scale the coordinates 
of a point freely and for any representative of \(p\in\TTAA^{d-1}\) all coordinates are finite,
this is possible for every coordinate. A common choice is to normalize the first coordinate, that is \[
    p = (p_1,p_2,\dots,p_d) \equiv (0, p_2-p_1, \dots, p_d - p_1)
\] where \(p_2-p_1, \dots, p_d-p_1\) constitute the coordinates of \(p\) in \(\RR^{d-1}\).

Under the identification of \(\TTAA^{d-1}\) with \(\RR^{d-1}\) we may compare tropical convexity
with classically convexity, but it turns out that these notions are not equivalent.
A tropical polytope is in general a polyhedral complex.
In the special case where a tropical polytope \(P\) is classically convex, we say that \(P\) is 
a \emph{polytrope} going back to \citeauthorandcite{JK:2010}.

Among tropical polytopes, polytropes enjoy many special properties that simplify computations.
For one, polytropes are tropical simplices, that is, for a polytrope \(P\subset\TTAA^{d-1}\)
we can find a square matrix \(C\in\TT^{d\times d}\) such that \(P=\tconv(C)\).
Additionally, polytropes are polyhedra in the classical sense which means they carry a facet description
given by classical linear inequalities. In particular, this facet description can be obtained as \[
    \wdp(C) \coloneqq \{\, x\in\TTAA^{d-1} \mid x_i - x_j \leq c_{ij} \text{ where } 1\leq i\neq j\leq d \,\}.
\] The construction of \(\wdp(C)\) is known as \emph{weighted digraph polyhedron}.

Our exposition suggests that there is a single tropical square matrix \(C\in\TT^{d\times d}\)
that realizes a given polytrope \(P\) as tropical convex hull \(\tconv(C)\) and as weighted digraph 
polyhedron \(\wdp(C)\). To make this connection precise, recall that the \emph{Kleene star} \(C^\star\)
of a tropical square matrix \(C\) is given by the power series \[
    C^\star \coloneqq I^n \oplus C \oplus C^2 \oplus \dots \oplus C^{d-1} \oplus \dots
\] where all operations are taken over the tropical semiring. 

Basic properties of the Kleene star are that multiplication by \(C^\star\) is idempotent, \(c^\star_{ii} = 0\) 
and that the entries satisfy the triangle inequality \begin{equation}
    c^\star_{ij} \leq c^\star_{ik} + c^\star_{kj}.
\end{equation}

Also, taking the Kleene star of a matrix \(C\) preserves weighted digraph polyhedra, 
that is \(\wdp(C) = \wdp(C^\star)\).
\begin{lemma}[{\cite[Lemma 3.\,24.]{Joswig:ETC}}]\label{lem:wdp-kleene-star}
    For any \(C\in\TT^{ \times d}\), if \(C^\star\) converges, then \(\wdp(C) = \wdp(C^\star)\).
\end{lemma}
In addition, the inequalities of \(\wdp(C^\star)\) are tight while some inequalities might be redundant.
In total, we can say that for a polytrope \(P\subseteq\TTAA^{d-1}\) there exists a
tropical square matrix \(C\in\TT^{d\times d}\) such that \begin{equation}\label{eq:presentation-of-polytrope}
    P = \tconv(C^\star) = \wdp(C^\star) = \wdp(C).
\end{equation} In particular, \(C^\star\) gives a tropical vertex and a classical facet description of \(P\).

\subsection{Max-linear Bayesian models and data generation}
We turn to max-linear Bayesian networks which are a \emph{recursive structural equation model}
defined over the max-times semiring and have been introduced by \citeauthor{GK:2018}~\cite{GK:2018}.

\begin{definition}
For a weighted DAG \(G = (V,E)\) with weight matrix \(C\in\RR^{d\times d}_{\geq 0}\) define the 
\emph{max-linear Bayesian network} (MLBN) \(X = (X_1,\dots, X_n)\) by the equations
\begin{equation}\label{eq:recursive-eqns}
    X_i = \bigvee_{i=1}^d c_{ij} X_j \vee c_{ii} Z_i, \quad i = 1, \dots, n,
\end{equation} where $Z=(Z_1,\dots,Z_n)$ are assumed to be independent random variables, each with support 
\(\RR_{>0}\) and atom-free distributions.
\end{definition}
We may write above recursive linear system of equations compactly as the max-times matrix-vector product \[
    X = C\cdot X \vee Z.
\] By repeated substitution, this recursive equation system admits the solution 
\begin{equation}\label{eq:MLBN-solution}
    X = C^\star\cdot Z
\end{equation} where \(C^\star\) is the Kleene star over the max-times semiring.
By assumption, \(C\) is the weight matrix of a directed acyclic graph making \(C^*\) always well-defined.

After applying a logarithmic transformation, the set of possible observations for \(X\) forms a polytrope in 
\(\TTAA^{d-1}\) due to \eqref{eq:presentation-of-polytrope} and \eqref{eq:MLBN-solution}.
This has two implications for MLBNs. For one, there is a correspondence between the weights of the underlying DAG
of an MLBN and the facets of \(\wdp(-\log{C})\). On the other hand, to generate data for simulations as in \Cref{sec:simulation},
it suffices to generate data for the innovations \(Z_i\) and (tropically) multiply those with the Kleene star of a 
weight matrix \(C\).

To make use of the identification of \(\TTAA^{d-1}\) with Euclidean space, we will implicitly apply a negative logarithmic 
transformation in the remainder of this work and consider MLBNs over the min-plus semiring.

\subsection{Regular subdivisions of root polytopes}
To study the combinatorial properties of polytropes, we define a certain point configuration as a combinatorial
dual. This has been subject in \cite{AF:2024,JS:2019} and we recall the necessary details in the following.

\begin{definition}
  For a digraph \(\thedag\) define the associated \emph{root polytope} as 
  \[
    \fundamentalpolytope[\thedag] 
    = \conv\left(
      \{0\} \cup \left\{\, e_{ij} \coloneqq e_i - e_j \mid j\to i\in\thedag \,\right\}
    \right)\subset\RR^d. \qedhere
  \]
\end{definition}

A polyhedral subdivision of a polytope \(P = \conv(V)\) in \(\RR^d\) is called \emph{regular} if it is
induced by a height function \(h\in\RR^V\). For further details, see \cite[Section A.\,4.]{Joswig:ETC}.
We say that a subdivision \(\Sigma\) of \(\fundamentalpolytope\) is \emph{central} if the vertex \(0\) is contained 
in every cell of \(\Sigma\).

Central subdivisions of root polytopes play an important role in the combinatorial characterization of polytropes
due to the following result that has been shown first by \citeauthorandcite{JS:2019} for the case of 
the complete graph \(\thedag = K_d\) and later by \citeauthorandcite{AF:2024} for arbitrary directed graphs.

\begin{theorem}[{\cite[Thm. 3.9]{AF:2024}}]\label{thm:central-triangulations}
    The combinatorial types of full-dimensional polytropes in \(\TTAA^{d-1}\) are in
    bijection with the regular central subdivisions of \(\fundamentalpolytope\) where \(\thedag\)
    ranges over transitive directed graphs \(\thedag\) on \(d\) nodes.
\end{theorem}

Since all subdivisions of interest are central and the origin is not an interior point of \(\fundamentalpolytope\)
whenever \(\thedag\) is acyclic, we may perform a small reduction step that allows us to reduce the dimensionality 
of the objects handled. In particular, for a central subdivision \(\Sigma\) of \(\fundamentalpolytope\), we may see every
cell \(\sigma\in\Sigma\) as a pyramid over a polytope only involving \(e_i-e_j\) vertices.

Thus, to any subdivision \(\Sigma\) of \(\fundamentalpolytope\) we can associate a subdivision \(\Sigma'\) of the vertices of 
\(\fundamentalpolytope\) excluding the origin \(0\) by taking for each cell \(\sigma\in\Sigma\) the unique facet not containing
the origin. This is called the \emph{link} of the origin in \(\Sigma\). 
An example of this is shown in \Cref{fig:link-of-subdivision}.

\begin{figure}[b]
    \centering
    \begin{subfigure}[c]{.45\textwidth}
        \centering
        \begin{tikzpicture}[
    roundnode/.style={circle, draw=black!85, fill=black!85, thin, scale=0.5},
    pseudonode/.style={circle, draw=black!85, fill=white, scale=0.5},
    sample/.style={draw=blue!85, fill=blue!85, thin},
    x  = {(0.9cm,-0.076cm)},
    y  = {(-0.06cm,0.95cm)},
    z  = {(-0.44cm,-0.29cm)},
    scale = 1,
    color = {lightgray}
]

  \coordinate (v0_unnamed__3) at (0, 0, 0);
  \coordinate (v1_unnamed__3) at (0, 0, 1);
  \coordinate (v2_unnamed__3) at (-1, 0, 1);
  \coordinate (v3_unnamed__3) at (0, -1, 1);

  \definecolor{vertexcolor_unnamed__3}{rgb}{ 1 0 0 }

  \tikzstyle{vertexstyle_unnamed__3_0} = [circle, scale=0.25, fill=vertexcolor_unnamed__3,label={[text=black, above right, align=left]:0},]
  \tikzstyle{vertexstyle_unnamed__3_1} = []
  \tikzstyle{vertexstyle_unnamed__3_2} = []
  \tikzstyle{vertexstyle_unnamed__3_3} = []

  \definecolor{facetcolor_unnamed__3}{rgb}{ 0.4666666667 0.9254901961 0.6196078431 }

  \definecolor{edgecolor_unnamed__3}{rgb}{ 0 0 0 }
  \tikzstyle{facetstyle_unnamed__3} = [fill=black, fill opacity=0.1, draw=edgecolor_unnamed__3, line width=1 pt, line cap=round, line join=round]

  \draw[facetstyle_unnamed__3] (v3_unnamed__3) -- (v2_unnamed__3) -- (v0_unnamed__3) -- (v3_unnamed__3) -- cycle;
  \draw[facetstyle_unnamed__3] (v1_unnamed__3) -- (v2_unnamed__3) -- (v3_unnamed__3) -- (v1_unnamed__3) -- cycle;
  \draw[facetstyle_unnamed__3] (v0_unnamed__3) -- (v2_unnamed__3) -- (v1_unnamed__3) -- (v0_unnamed__3) -- cycle;
  \draw[facetstyle_unnamed__3] (v3_unnamed__3) -- (v0_unnamed__3) -- (v1_unnamed__3) -- (v3_unnamed__3) -- cycle;

  \foreach \i in {2,3,1,0} {
    \node at (v\i_unnamed__3) [vertexstyle_unnamed__3_\i] {};
  }

  \coordinate (v0_unnamed__2) at (0, 0, 0);
  \coordinate (v1_unnamed__2) at (0, 0, 1);
  \coordinate (v2_unnamed__2) at (-1, 1, 0);
  \coordinate (v3_unnamed__2) at (-1, 0, 1);

  \definecolor{vertexcolor_unnamed__2}{rgb}{ 1 0 0 }

  \tikzstyle{vertexstyle_unnamed__2_0} = [circle, scale=0.25, fill=vertexcolor_unnamed__2,label={[text=black, above right, align=left]:0},]
  \tikzstyle{vertexstyle_unnamed__2_1} = []
  \tikzstyle{vertexstyle_unnamed__2_2} = []
  \tikzstyle{vertexstyle_unnamed__2_3} = [circle, scale=0.25, fill=vertexcolor_unnamed__2,label={[text=black, above left, align=left]:$e_{42}$},]

  \definecolor{facetcolor_unnamed__2}{rgb}{ 0.4666666667 0.9254901961 0.6196078431 }

  \definecolor{edgecolor_unnamed__2}{rgb}{ 0 0 0 }
  \tikzstyle{facetstyle_unnamed__2} = [fill=black, fill opacity=0.1, draw=edgecolor_unnamed__2, line width=1 pt, line cap=round, line join=round]

  \draw[facetstyle_unnamed__2] (v3_unnamed__2) -- (v2_unnamed__2) -- (v0_unnamed__2) -- (v3_unnamed__2) -- cycle;
  \draw[facetstyle_unnamed__2] (v3_unnamed__2) -- (v0_unnamed__2) -- (v1_unnamed__2) -- (v3_unnamed__2) -- cycle;
  \draw[facetstyle_unnamed__2] (v1_unnamed__2) -- (v2_unnamed__2) -- (v3_unnamed__2) -- (v1_unnamed__2) -- cycle;
  \draw[facetstyle_unnamed__2] (v0_unnamed__2) -- (v2_unnamed__2) -- (v1_unnamed__2) -- (v0_unnamed__2) -- cycle;

  \foreach \i in {3,1,2,0} {
    \node at (v\i_unnamed__2) [vertexstyle_unnamed__2_\i] {};
  }
  
  \coordinate (v0_unnamed__1) at (0, 0, 0);
  \coordinate (v1_unnamed__1) at (0, 1, 0);
  \coordinate (v2_unnamed__1) at (0, 0, 1);
  \coordinate (v3_unnamed__1) at (-1, 1, 0);

  \definecolor{vertexcolor_unnamed__1}{rgb}{ 1 0 0 }

  \tikzstyle{vertexstyle_unnamed__1_0} = [circle, scale=0.25, fill=vertexcolor_unnamed__1,label={[text=black, above right, align=left]:0},]
  \tikzstyle{vertexstyle_unnamed__1_1} = []
  \tikzstyle{vertexstyle_unnamed__1_2} = []
  \tikzstyle{vertexstyle_unnamed__1_3} = [circle, scale=0.25, fill=vertexcolor_unnamed__1,label={[text=black, above left, align=left]:$e_{32}$},]

  \definecolor{facetcolor_unnamed__1}{rgb}{ 0.4666666667 0.9254901961 0.6196078431 }

  \definecolor{edgecolor_unnamed__1}{rgb}{ 0 0 0 }
  \tikzstyle{facetstyle_unnamed__1} = [fill=black, fill opacity=0.1, draw=edgecolor_unnamed__1, line width=1 pt, line cap=round, line join=round]

  \draw[facetstyle_unnamed__1] (v3_unnamed__1) -- (v0_unnamed__1) -- (v2_unnamed__1) -- (v3_unnamed__1) -- cycle;
  \draw[facetstyle_unnamed__1] (v1_unnamed__1) -- (v0_unnamed__1) -- (v3_unnamed__1) -- (v1_unnamed__1) -- cycle;
  \draw[facetstyle_unnamed__1] (v3_unnamed__1) -- (v2_unnamed__1) -- (v1_unnamed__1) -- (v3_unnamed__1) -- cycle;
  \draw[facetstyle_unnamed__1] (v2_unnamed__1) -- (v0_unnamed__1) -- (v1_unnamed__1) -- (v2_unnamed__1) -- cycle;

  \foreach \i in {2,3,0,1} {
    \node at (v\i_unnamed__1) [vertexstyle_unnamed__1_\i] {};
  }
  
  \coordinate (v0_unnamed__5) at (0, 0, 0);
  \coordinate (v1_unnamed__5) at (1, 0, 0);
  \coordinate (v2_unnamed__5) at (0, 0, 1);
  \coordinate (v3_unnamed__5) at (0, -1, 1);

  \definecolor{vertexcolor_unnamed__5}{rgb}{ 1 0 0 }

  \tikzstyle{vertexstyle_unnamed__5_0} = [circle, scale=0.25, fill=vertexcolor_unnamed__5,label={[text=black, above right, align=left]:0},]
  \tikzstyle{vertexstyle_unnamed__5_1} = []
  \tikzstyle{vertexstyle_unnamed__5_2} = []
  \tikzstyle{vertexstyle_unnamed__5_3} = [circle, scale=0.25, fill=vertexcolor_unnamed__5,label={[text=black, below right, align=left]:$e_{43}$},]

  \definecolor{facetcolor_unnamed__5}{rgb}{ 0.4666666667 0.9254901961 0.6196078431 }

  \definecolor{edgecolor_unnamed__5}{rgb}{ 0 0 0 }
  \tikzstyle{facetstyle_unnamed__5} = [fill=black, fill opacity=0.1, draw=edgecolor_unnamed__5, line width=1 pt, line cap=round, line join=round]

  \draw[facetstyle_unnamed__5] (v0_unnamed__5) -- (v3_unnamed__5) -- (v2_unnamed__5) -- (v0_unnamed__5) -- cycle;
  \draw[facetstyle_unnamed__5] (v1_unnamed__5) -- (v3_unnamed__5) -- (v0_unnamed__5) -- (v1_unnamed__5) -- cycle;
  \draw[facetstyle_unnamed__5] (v2_unnamed__5) -- (v3_unnamed__5) -- (v1_unnamed__5) -- (v2_unnamed__5) -- cycle;
  \draw[facetstyle_unnamed__5] (v0_unnamed__5) -- (v2_unnamed__5) -- (v1_unnamed__5) -- (v0_unnamed__5) -- cycle;

  \foreach \i in {3,2,0,1} {
    \node at (v\i_unnamed__5) [vertexstyle_unnamed__5_\i] {};
  }

  \coordinate (v0_unnamed__4) at (0, 0, 0);
  \coordinate (v1_unnamed__4) at (1, 0, 0);
  \coordinate (v2_unnamed__4) at (0, 1, 0);
  \coordinate (v3_unnamed__4) at (0, 0, 1);

  \definecolor{vertexcolor_unnamed__4}{rgb}{ 1 0 0 }

  \tikzstyle{vertexstyle_unnamed__4_0} = [circle, scale=0.25, fill=vertexcolor_unnamed__4,label={[text=black, above right, align=left]:0},]
  \tikzstyle{vertexstyle_unnamed__4_1} = [circle, scale=0.25, fill=vertexcolor_unnamed__4,label={[text=black, above right, align=left]:$e_{21}$},]
  \tikzstyle{vertexstyle_unnamed__4_2} = [circle, scale=0.25, fill=vertexcolor_unnamed__4,label={[text=black, above right, align=left]:$e_{31}$},]
  \tikzstyle{vertexstyle_unnamed__4_3} = [circle, scale=0.25, fill=vertexcolor_unnamed__4,label={[text=black, below left, align=left]:$e_{41}$},]

  \definecolor{facetcolor_unnamed__4}{rgb}{ 0.4666666667 0.9254901961 0.6196078431 }

  \definecolor{edgecolor_unnamed__4}{rgb}{ 0 0 0 }
  \tikzstyle{facetstyle_unnamed__4} = [fill=black, fill opacity=0.1, draw=edgecolor_unnamed__4, line width=1 pt, line cap=round, line join=round]

  \draw[facetstyle_unnamed__4] (v0_unnamed__4) -- (v3_unnamed__4) -- (v2_unnamed__4) -- (v0_unnamed__4) -- cycle;
  \draw[facetstyle_unnamed__4] (v1_unnamed__4) -- (v3_unnamed__4) -- (v0_unnamed__4) -- (v1_unnamed__4) -- cycle;
  \draw[facetstyle_unnamed__4] (v0_unnamed__4) -- (v2_unnamed__4) -- (v1_unnamed__4) -- (v0_unnamed__4) -- cycle;

   \node at (v0_unnamed__4) [vertexstyle_unnamed__4_0] {};

  \draw[facetstyle_unnamed__4] (v2_unnamed__4) -- (v3_unnamed__4) -- (v1_unnamed__4) -- (v2_unnamed__4) -- cycle;

  \foreach \i in {3,2,1} {
    \node at (v\i_unnamed__4) [vertexstyle_unnamed__4_\i] {};
  }

\end{tikzpicture}
        \caption{Full central subdivision}
    \end{subfigure}
    \begin{subfigure}[c]{.45\textwidth}
        \centering
        \begin{tikzpicture}[
    roundnode/.style={circle, draw=black!85, fill=black!85, thin, scale=0.5},
    pseudonode/.style={circle, draw=black!85, fill=white, scale=0.5},
    sample/.style={draw=blue!85, fill=blue!85, thin},
    scale = 1,
    color = {lightgray}
]

  \coordinate (v0_unnamed__1) at (0, 1);
  \coordinate (v1_unnamed__1) at (0, 0);
  \coordinate (v2_unnamed__1) at (-1, 1);

  \definecolor{vertexcolor_unnamed__1}{rgb}{ 1 0 0 }

  \tikzstyle{vertexstyle_unnamed__1_0} = []
  \tikzstyle{vertexstyle_unnamed__1_1} = []
  \tikzstyle{vertexstyle_unnamed__1_2} = []

  \definecolor{edgecolor_unnamed__1}{rgb}{ 0 0 0 }
  \tikzstyle{facetstyle_unnamed__1} = [fill=black, fill opacity=0.15, draw=edgecolor_unnamed__1, line width=.5 pt, line cap=round, line join=round]

  \draw[facetstyle_unnamed__1] (v1_unnamed__1) -- (v0_unnamed__1) -- (v2_unnamed__1) -- (v1_unnamed__1) -- cycle;

  \foreach \i in {0,1,2} {
    \node at (v\i_unnamed__1) [vertexstyle_unnamed__1_\i] {};
  }

  \coordinate (v0_unnamed__2) at (0, 0);
  \coordinate (v1_unnamed__2) at (-1, 1);
  \coordinate (v2_unnamed__2) at (-1, 0);

  \definecolor{vertexcolor_unnamed__2}{rgb}{ 1 0 0 }

  \tikzstyle{vertexstyle_unnamed__2_0} = []
  \tikzstyle{vertexstyle_unnamed__2_1} = [circle, scale=0.4, fill=black,label={[text=black, above left, align=right]:$\scriptstyle e_{32}$},]
  \tikzstyle{vertexstyle_unnamed__2_2} = []
  
  \definecolor{edgecolor_unnamed__2}{rgb}{ 0 0 0 }
  \tikzstyle{facetstyle_unnamed__2} = [fill=black, fill opacity=0.15, draw=edgecolor_unnamed__2, line width=.5 pt, line cap=round, line join=round]

  \draw[facetstyle_unnamed__2] (v2_unnamed__2) -- (v0_unnamed__2) -- (v1_unnamed__2) -- (v2_unnamed__2) -- cycle;

  \foreach \i in {0,1,2} {
    \node at (v\i_unnamed__2) [vertexstyle_unnamed__2_\i] {};
  }

  \coordinate (v0_unnamed__3) at (0, 0);
  \coordinate (v1_unnamed__3) at (-1, 0);
  \coordinate (v2_unnamed__3) at (0, -1);

  \definecolor{vertexcolor_unnamed__3}{rgb}{ 1 0 0 }

  \tikzstyle{vertexstyle_unnamed__3_0} = []
  \tikzstyle{vertexstyle_unnamed__3_1} = [circle, scale=0.4, fill=black,label={[text=black, above left, align=left]:$\scriptstyle e_{42}$},]
  \tikzstyle{vertexstyle_unnamed__3_2} = []

  \definecolor{edgecolor_unnamed__3}{rgb}{ 0 0 0 }
  \tikzstyle{facetstyle_unnamed__3} = [fill=black, fill opacity=0.15, draw=edgecolor_unnamed__3, line width=.5 pt, line cap=round, line join=round]

  \draw[facetstyle_unnamed__3] (v2_unnamed__3) -- (v1_unnamed__3) -- (v0_unnamed__3) -- (v2_unnamed__3) -- cycle;

  \foreach \i in {0,1,2} {
    \node at (v\i_unnamed__3) [vertexstyle_unnamed__3_\i] {};
  }

  \coordinate (v0_unnamed__5) at (1, 0);
  \coordinate (v1_unnamed__5) at (0, 0);
  \coordinate (v2_unnamed__5) at (0, -1);

  \definecolor{vertexcolor_unnamed__5}{rgb}{ 1 0 0 }

  \tikzstyle{vertexstyle_unnamed__5_0} = [circle, scale=0.4, fill=black,label={[text=black, above right, align=left]:$\scriptstyle e_{21}$},]
  \tikzstyle{vertexstyle_unnamed__5_1} = [circle, scale=0.4, fill=black,label={[text=black, below right, align=left]:$\scriptstyle e_{41}$},]
  \tikzstyle{vertexstyle_unnamed__5_2} = [circle, scale=0.4, fill=black,label={[text=black, below right, align=left]:$\scriptstyle e_{43}$},]

  \definecolor{edgecolor_unnamed__5}{rgb}{ 0 0 0 }
  \tikzstyle{facetstyle_unnamed__5} = [fill=black, fill opacity=0.15, draw=edgecolor_unnamed__5, line width=.5 pt, line cap=round, line join=round]

  \draw[facetstyle_unnamed__5] (v2_unnamed__5) -- (v0_unnamed__5) -- (v1_unnamed__5) -- (v2_unnamed__5) -- cycle;

  \foreach \i in {0,1,2} {
    \node at (v\i_unnamed__5) [vertexstyle_unnamed__5_\i] {};
  }

  \coordinate (v0_unnamed__4) at (1, 0);
  \coordinate (v1_unnamed__4) at (0, 1);
  \coordinate (v2_unnamed__4) at (0, 0);

  \definecolor{vertexcolor_unnamed__4}{rgb}{ 1 0 0 }

  \tikzstyle{vertexstyle_unnamed__4_0} = []
  \tikzstyle{vertexstyle_unnamed__4_1} = [circle, scale=0.4, fill=black,label={[text=black, above right, align=left]:$\scriptstyle e_{31}$},]
  \tikzstyle{vertexstyle_unnamed__4_2} = []

  \definecolor{edgecolor_unnamed__4}{rgb}{ 0 0 0 }
  \tikzstyle{facetstyle_unnamed__4} = [fill=black, fill opacity=0.15, draw=black, line width=.5 pt, line cap=round, line join=round]

  \draw[facetstyle_unnamed__4] (v2_unnamed__4) -- (v0_unnamed__4) -- (v1_unnamed__4) -- (v2_unnamed__4) -- cycle;

  \foreach \i in {0,1,2} {
    \node at (v\i_unnamed__4) [vertexstyle_unnamed__4_\i] {};
  }

\end{tikzpicture}
        \caption{Link of \(0\)}
    \end{subfigure}
    \caption{A central triangulation \(\Sigma\) of the root polytope \(\fundamentalpolytope[\kappa_4]\) and the link 
    of the origin in \(\Sigma\).}
    \label{fig:link-of-subdivision}
\end{figure}

We say that a MLBN supported on a DAG \(\thedag\) is \emph{generic} if the weight matrix \(C\) induces a central 
triangulation of \(\fundamentalpolytope\). The subdivision of \(\fundamentalpolytope\) induced in this way
is called the \emph{dual subdivision} to \(\wdp(C)\).

This correspondence has a consequence on the number of pseudovertices of a polytrope. This is stated in the following
well-known result that has been noted by \citeauthorandcite{DS:2004} for tropical polytopes but ultimately goes back 
to work of \citeauthorandcite{GGP:1997}. We include a short summary of the proof.
\begin{lemma}\label{lem:catalan-number-pseudovertices}
    If \(C\in\TT^{d\times d}\) is a lower-triangular matrix and \(\wdp(C)\) is full-dimensional, 
    the number of pseudovertices of \(\wdp(C)\) is at most the \((d-1)\)-st Catalan number 
    \[
      C_{d-1} = \frac{1}{d}\binom{2(d-1)}{d-1}.
    \]
    \begin{proof}
        A polytrope \(\wdp(C)\) for lower-triangular \(C\) is dual to a subdivision of a 
        subpolytope of \(P_{\kappa_d}\). The root polytope \(P_{\kappa_d}\) itself has
        been studied by \citeauthor{GGP:1997} and has been found to have normalized volume
        equal to the \((d-1)\)-st Catalan number \(C_{d-1}\) \cite[Theorem 2.3]{GGP:1997}.
        In particular, a triangulation of \(P_{\kappa_d}\) has \(C_{d-1}\) many simplices
        and \(\wdp(C)\) has as many pseudovertices.
    \end{proof}
\end{lemma}

\section{Calculating minimum bounding polytropes}\label{sec:minimum-bounding-polytrope}
The goal of this section is to calculate a \emph{minimum bounding polytrope} 
\(P(S)\subset\TTAA^{d-1}\) efficiently given a finite set of points \(S = \{\, p^{(1)}, \dots, p^{(n)}\}\subset\TTAA^{d-1}\).
We can simplify this problem by making use of the classical facet description of a polytrope
given by inequalities \begin{equation}\label{eq:polytrope-facets}
    x_i - x_j \leq c_{ij}
\end{equation} for a matrix \(C\in\TT^{d\times d}\) with \(c_{ii} = 0\).

We obtain a matrix \(\widetilde{C}\in\TT^{d\times d}\) by taking all maxima over \(S\) 
of the coordinate differences, \begin{equation}\label{eq:minimum-bounding-facets}
    \widetilde{c}_{ij} \coloneqq \max_kp^{(k)}_i - p^{(k)}_j.
\end{equation}
\begin{proposition}\label{prop:minimum-bounding-polytrope}
    Let \(S\subset\TTAA^{d-1}\) be a set of \(n\) points. Then, the matrix \(\widetilde{C}\) 
    provides a facet description of \(P(S)\) and the inequalities are tight. In particular,
    \(\widetilde{C}^\star = \widetilde{C}\).

    \begin{proof}
         The first part is by construction. To see that \(\widetilde{C}^\star = \widetilde{C}\),
         assume to the contrary. Then, there is an entry \(\widetilde{c}_{ij}^\star\lneq\widetilde{c}_{ij}\).
         There also exists a point \(p^{(k)}\) such that \(\widetilde{c}_{ij} = p_i^{(k)} - p_j^{(j)}\).
         But then, \(p^{(k)}\in\wdp(\widetilde{C})\) and \(p^{(k)}\notin\wdp(\widetilde{C}^\star)\),
         which means \(\wdp(\widetilde{C})\neq\wdp(\widetilde{C}^\star)\), contradicting \Cref{lem:wdp-kleene-star}.
    \end{proof}
\end{proposition}
This also shows that \(P(S)\) is a smallest polytrope with the property that \(S\subset P(S)\), 
and \(\widetilde{C}\) constitutes a tropical vertex description of \(P(S)\), which means
\({P(S) = \tconv(\widetilde{C})}\).

We also get an upper bound on the calculation of the minimum bounding polytrope from \Cref{prop:minimum-bounding-polytrope}.
This has been observed before by \citeauthorandcite[Lem.\ S2]{BKT:2024} in the context of estimating MLBNs on trees
and by \citeauthorandcite{Min:2001} in the context of static program analysis with loops over domains that
can be represented by polytropes.
\begin{proposition}
    For a finite set \(S\subset\TTAA^{d-1}\) of \(n\) points, the minimum bounding polytrope \(P(S)\) can be calculated 
    using \(\bigo(d^2n)\)-many comparison operations.
\end{proposition}

In the next step, we now assume that \(S\subset\wdp(C)\) for some matrix \(C\in\TT^{d\times d}\).
For this, we have the following expected containment.

\begin{lemma}
    Let \(S\subset\wdp(C)\) be a set of \(n\) points. Then, \(P(S)\subset\wdp(C)\).

    \begin{proof}
        It follows from \eqref{eq:polytrope-facets} that for every \(p\in S\) the inequalities \[
            p_i - p_j \leq c_{ij}
        \] are satisfied for all \(i\neq j\in [d]\). But this also gives the bound \[
            \widetilde{c}_{ij} = \max_kp^{(k)}_i - p^{(k)}_j \leq c_{ij}.
        \] Applying \eqref{eq:polytrope-facets} to \(P(S)\) implies that every point \(p\in P(S)\) satisfies \[
            p_i - p_j \leq \widetilde{c}_{ij} \leq c_{ij}
        \] which means that \(P(S)\subset\wdp(C)\).
    \end{proof}
\end{lemma}

Since the weight \(c_{ij}\) might not define a facet of \(\wdp(C)\), and for specific \(S\subset\wdp(C)\)
we might have \(\widetilde{c}_{ij}\lneq c_{ij}\), \(P(S)\) might exhibit more facets than \(\wdp(C)\).

\begin{figure}[t]
    \centering
    \begin{tikzpicture}[
    roundnode/.style={circle, draw=black!85, fill=black!85, thin, scale=0.5},
    pseudonode/.style={circle, draw=black!85, fill=white, scale=0.5},
    sample/.style={draw=blue!85, fill=blue!85, thin},
    scale=2
]            
    \coordinate (v1) at (-1,0);
    \coordinate (v2) at (-1,0);

    \coordinate (v3) at (-2,-1);
    \coordinate (v4) at (-1,-1.5);

    \draw[black!15, fill=black!15] (v1) -- (v2) -- (v3) --++ (0,-.5) --++ (1,0) -- cycle;

    \draw[black!85!black, line width=2pt] (v2) -- (v3); 
    \draw[black!85!black, line width=2pt] (v1) -- (v4);

    \node[pseudonode] at (v2) {};
    \node[roundnode] at (v1) {};

    \node[roundnode,blue!80] at (-1,-.5) {};
    \node[roundnode,blue!80] at (-1.5,-.5) {};
    \draw[blue!85, loosely dotted, line width=2pt] (-2,-.5) -- (-.5,-.5);

\end{tikzpicture}
    \caption{A schematic of \Cref{ex:underestimate-additional-facets} where \(P(S)\) exhibits 
    more facets than \(\wdp(C)\). The additional facet of \(P(S)\) is given by the dotted blue line.}
    \label{fig:spurious-facet}
\end{figure}
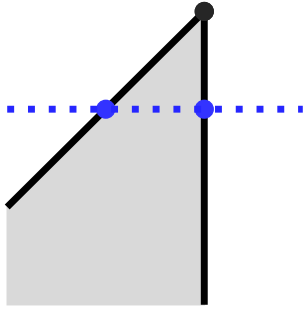
\begin{example}\label{ex:underestimate-additional-facets}
    Suppose \(\thedag = \kappa_3\) and \(C\in\TT^{3\times 3}\) such that \(c_{31}\geq c_{32} + c_{21}\).
    In this case, the \(x_3-x_1\)-hyperplane does not define a facet of \(\wdp(C)\). 
    If in a sample \(S\) we observe that \(\widetilde{c}_{31}<c^\star_{31}\), then in particular \(P(S)\)
    will have a facet supported by a \(x_3-x_1\)-hyperplane. This is shown in \Cref{fig:spurious-facet}.
\end{example}

\section{Parameter recovery and set covers}\label{sec:param-recovery}
In this section, we discuss estimating the \emph{true weight matrix} \(C\) of the MLBN using the minimum bounding polytrope
in the setting where the underlying DAG \(\thedag\) is known.
Since a sample \(S\) is necessarily bounded, the maxima of every difference \(X_{ij} \coloneqq X_i - X_j\) will be finite
and \(\widetilde{C}\) will be supported on the complete graph \(K_d\). 
Thus, estimating \(C\) from \(\widetilde{C}\) includes selecting a subset of the entries that give a matrix 
\(\estimateC\in\TT^{d\times d}\) that is supported on a DAG. In the current setting where the underlying DAG \(\thedag\) 
is known, we set \[
    \estimateC\coloneqq\widetilde{C}\vert_\thedag
    \quad\text{where}\quad
    \widehat{c}_{ij} \coloneqq\begin{cases}
        \widetilde{c}_{ij},& \text{if } j\to i\in\thedag,\\
        \infty,&\text{otherwise}.
    \end{cases}
\]

Note that since \(\wdp(C) = \wdp(C^\star)\), we also have \(c_{ij}^\star\leq c_{ij}\). 
Depending on the true weights, this makes it impossible to recover certain \(c_{ij}\), 
since we can at most expect to recover the entries of \(C^\star\).
The extent to which the parameters of an MLBN can be identified has been studied by \citeauthorandcite{GKL:2021}
while \citeauthorandcite{AF:2024} gave a polyhedral description of the parameter matrices that could be observed for fixed
true parameter matrix. For this reason, we assume that \(C^\star = C\) and that we are trying to estimate this Kleene star.

Note that the obtained matrix \(\estimateC\) is not necessarily a Kleene star which necessarily happens when the known
DAG \(\thedag\) is not transitive. Since \citeauthor{GKL:2021} specifically estimate the Kleene star of \(C\), 
they additionally took the Kleene star of their estimate.
Yet we decide to focus on the simpler estimator \(\estimateC\) without additionally taking the Kleene star of \(\estimateC\).
Since in the case where \(\thedag\) is known, we are only interested in the edge weights of \(\thedag\) and not in
the redundant information provided by the Kleene star.

In addition, \citeauthorandcite{GKL:2021} showed that the distribution of the difference \(X_{ij}\) 
contains atoms, one of which precisely occurs at \(c_{ij}\). 
As a result, there is a positive probability that there is a sample point \(X^{(k)}\in S\) for which \(X^{(k)}_{ij} = c_{ij}\).

For these reasons, in any sample \(S\subset\wdp(C)\) only the presence of special points is necessary for
\(\estimateC\) to recover the correct parameters. A natural question to ask is how many of these special points are 
necessary in the best case to recover \(C\) from \(\estimateC\).

\begin{figure}[t]
    \centering
    \begin{subfigure}[c]{0.65\textwidth}
        \begin{tikzpicture}[
    roundnode/.style={circle, draw=black!85, fill=black!85, thin, scale=0.5},
    pseudonode/.style={circle, draw=black!85, fill=white, scale=0.5},
    sample/.style={draw=blue!85, fill=blue!85, thin},
]

    \coordinate (0) at (0,0);
    \coordinate (12) at (-1,1,0);
    \coordinate (21) at (1,-1,0);
    \coordinate (13) at (-1,0,1);
    \coordinate (31) at (1,0,-1);
    \coordinate (23) at (0,-1,1);
    \coordinate (32) at (0,1,-1);
    
    \draw[orange!85,fill=orange!15,very thick] (0) -- (12) -- (13) -- (23) -- cycle;
    \draw[black] (0) -- (13);

    \draw[blue!85,fill=blue!15,very thick] (0) -- (13) -- (23) -- cycle;
    \draw[orange!85,very thick] (0) -- (12);

    \node[roundnode] at (0) {};
    \node[roundnode,orange!85] at (12) {}; \node[anchor=south east] at (12) {$e_{21}$};
    \node[roundnode] at (21) {};
    \node[roundnode,blue!85] at (13) {}; \node[roundnode,orange!85,scale=.5] at (13) {}; 
    \node[anchor=east] at (13) {$e_{31}$};
    \node[roundnode] at (31) {}; 
    \node[roundnode,blue!85] at (23) {}; \node[anchor=north] at (23) {$e_{32}$};
    \node[roundnode] at (32) {};

    \begin{scope}[xshift=7cm,yshift=2em,scale=2]
        \node at (.5,.5) {};
        \node at (0,-1) {};
    
        \coordinate (v1) at (0,0);
        \coordinate (v2) at (-1,0);
    
        \coordinate (v3) at (-2,-1);
        \coordinate (v4) at (0,-1);
    
        \coordinate (u1) at (.5,0);
        \coordinate (u2) at (-2,-1);
    
        \coordinate (u3) at (0,.5);
    
        \draw[black!15, fill=black!15] (v1) -- (v2) -- (v3) --++ (2,0) -- cycle;
    
        \draw[black!85!black, line width=2pt] (v2) -- (v3); 
        \draw[black!85!black, line width=2pt] (v2) -- (v1);
        \draw[black!85!black, line width=2pt] (v1) -- node[near end,inner sep=0] (s0) {}  (v4);
    
        \node[roundnode] at (v1) {};
        \node[pseudonode] at (v2) {};

        \node[sample,orange] at (v1) {};
        \node[sample] at (v2) {};
    \end{scope}
\end{tikzpicture}
        \caption{Sample \(S_1\) given by pseudovertices.}
        \label{fig:covering-pv}
    \end{subfigure}
    \begin{subfigure}[c]{0.65\textwidth}
        \begin{tikzpicture}[
    roundnode/.style={circle, draw=black!85, fill=black!85, thin, scale=0.5},
    pseudonode/.style={circle, draw=black!85, fill=white, scale=0.5},
    sample/.style={draw=blue!85, fill=blue!85, thin},
]
    \coordinate (0) at (0,0);
    \coordinate (12) at (-1,1,0);
    \coordinate (21) at (1,-1,0);
    \coordinate (13) at (-1,0,1);
    \coordinate (31) at (1,0,-1);
    \coordinate (23) at (0,-1,1);
    \coordinate (32) at (0,1,-1);
    
    \draw[black, fill=black!15,thick] (0) -- (12) -- (13) -- (23) -- cycle;
    \draw[black,fill=black!15,thick] (0) -- (13) -- (23) -- cycle;

    \node[roundnode] at (0) {};
    \node[roundnode,blue!85] at (12) {}; \node[anchor=south east] at (12) {$e_{21}$};
    \node[roundnode] at (21) {};
    \node[roundnode,orange!85] at (13) {}; \node[anchor=east] at (13) {$e_{31}$};
    \node[roundnode] at (31) {}; 
    \node[roundnode,magenta!85] at (23) {}; \node[anchor=north] at (23) {$e_{32}$};
    \node[roundnode] at (32) {};

    \begin{scope}[xshift=7cm,yshift=2em,scale=2]
        \node at (.5,.5) {};
        \node at (0,-1) {};
    
        \coordinate (v1) at (0,0);
        \coordinate (v2) at (-1,0);
    
        \coordinate (v3) at (-2,-1);
        \coordinate (v4) at (0,-1);
    
        \coordinate (u1) at (.5,0);
        \coordinate (u2) at (-2,-1);
    
        \coordinate (u3) at (0,.5);
    
        \draw[black!15, fill=black!15] (v1) -- (v2) -- (v3) --++ (2,0) -- cycle;
    
        \draw[black!85!black, line width=2pt] (v2) -- (v3); 
        \draw[black!85!black, line width=2pt] (v2) -- (v1);
        \draw[black!85!black, line width=2pt] (v1) -- node[near end,inner sep=0] (s0) {}  (v4);
    
        \node[roundnode] at (v1) {};
        \node[pseudonode] at (v2) {};

        \node[sample,orange] at (-.5,0) {};
        \node[sample,magenta] at (-1.5,-.5) {};
        \node[sample] at (0,-.5) {};
    \end{scope}
\end{tikzpicture}
        \caption{Sample \(S_2\) given by points on the relative interior of the facets.}
        \label{fig:covering-facets}
    \end{subfigure}
    \caption{Two examples of \(\widetilde\Sigma\) for samples from \(\wdp(C)\) in \Cref{ex:dual-covering}.}
    \label{fig:dual-covering}
\end{figure}

For this, let \(\Sigma\) be the link of the origin of the regular subdivision of \(\fundamentalpolytope\) 
dual to \(\wdp(C)\) and define the following map from \(\wdp(C)\) to \(\Sigma\): 
associate to each point \(p\in\wdp(C)\) the cell \[
    \{\,
      e_{ji} \mid p_i-p_j = c_{ij}
    \,\} 
\] of the subdivision of \(\fundamentalpolytope\). Then, denote by \(\widetilde\Sigma(S)\) the set of cells for each 
\(p\in S\).
The question whether \(\widetilde{C} = C\) for a given sample \(S\subset\wdp(C)\) is then equivalent
to asking whether for every point \(e_{ji}\in\fundamentalpolytope\) there is a \(\sigma\in\widetilde\Sigma(S)\) 
containing that point. That is, we ask for \(\widetilde\Sigma(S)\) to be a set cover for the subdivision \(\Sigma\).
We demonstrate this with the following example.

\begin{example}\label{ex:dual-covering}
    Consider the MLBN given by the weight matrix \[
        C = \begin{pmatrix}
            0 & \infty & \infty \\
            1 & 0 & \infty \\
            2 & 3 & 0
        \end{pmatrix}
    \] and take the two samples \begin{align*}
        S_1 &= \{\,
            (0,-1,2), (0,1,1) 
        \,\} \\
        S_2 &= \{\,
            (0,0,2), (0,1,1), (0,0,1)
        \,\}.
    \end{align*}
    The sets \(\widetilde\Sigma\) for both samples are shown in \Cref{fig:dual-covering}; they are \begin{align*}
        \widetilde\Sigma(S_1) &= \left\{\,
          \{ e_{12} \}, \{ e_{13}, e_{23} \}
        \,\right\},\text{ and} \\
        \widetilde\Sigma(S_2) &= \{\,
          \{ e_{13} \}, \{ e_{23} \}, \{ e_{12} \}
        \,\}.
    \end{align*}
    It can be seen in \Cref{fig:dual-covering} that \(S_1\) covers every facet of \(\wdp(C)\) 
    using two points while \(S_2\) covers every facet in three points.
\end{example}

Above example also shows two natural choices for special points in samples for recovering \(C\) from
\(\estimateC\) which correspond to two trivial set covers of \(\Sigma\).

If a sample \(S\) contains a point in each facet as in \Cref{fig:covering-facets}, 
\(\widetilde\Sigma(S)\) certainly covers the vertices of \(\Sigma\). This leaves us with a sample of
size at most \(\frac{(d-1)(d-2)}{2}\), since this is the maximal number of facets for \(\wdp(C)\) when \(C\)
is supported on a DAG.

On the other hand, if \(\widetilde\Sigma(S)\) is the set of all maximal dimensional cells of \(\Sigma\)
we also get a set cover. In this case, \(S\) has to contain all the pseudovertices of \(\wdp(C)\)
of which there are as many as the \((d-1)\)-st Catalan number \(C_{d-1}\) by 
\Cref{lem:catalan-number-pseudovertices}.

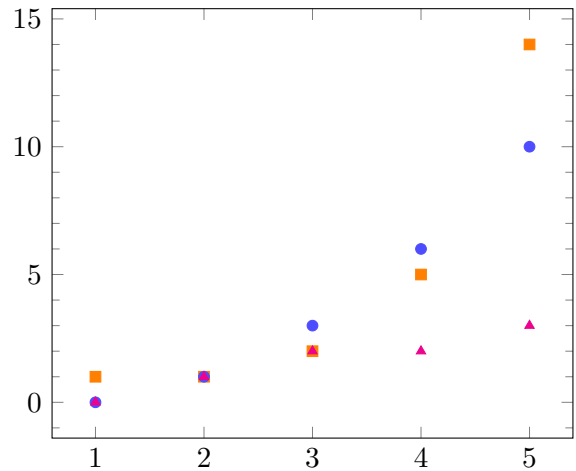
\begin{figure}[bt]
    \centering








\begin{tikzpicture}
\begin{axis}[enlargelimits, minor y tick num=4]
\addplot [
    scatter,
    mark=square*,
    only marks,
    scatter/use mapped color={
        fill=orange,
        draw=orange
    }
] table[meta=label] {
x y label
1 1 pseudovertices
2 1 pseudovertices
3 2 pseudovertices
4 5 pseudovertices
5 14 pseudovertices
};
    \addplot [
        scatter,
        mark=*,
        only marks,
        scatter/use mapped color={
            fill=blue!70,
            draw=blue!70
        }
    ] table[meta=label] {
x y label
1 0 facets
2 1 facets
3 3 facets
4 6 facets
5 10 facets
};

    \addplot [
        scatter,
        mark=triangle*,
        only marks,
        scatter/use mapped color={
            fill=magenta,
            draw=magenta
        }
    ] table[meta=label] {
x y label
1 0 minimal sample size
2 1 minimal sample size
3 2 minimal sample size
4 2 minimal sample size
5 3 minimal sample size
};

\end{axis}
\end{tikzpicture}
    \caption{The maximal number of facets (blue circles) and pseudovertices (orange squares)
    of a polytrope \(\wdp(C)\subset\TTAA^{d-1}\) when \(C\) is supported on a DAG. Also shown is the minimal best case sample size
    (magenta triangles) when \(C\) is supported on \(\kappa_n\) (compare with \Cref{tab:enumeration}).}
    \label{fig:facets-vs-pseudovertices-vs-actual-minimal-sample-size}
\end{figure}

\Cref{fig:facets-vs-pseudovertices-vs-actual-minimal-sample-size} shows a comparison of the maximal number of facets 
with the maximal number of pseudovertices for a \(d-1\)-dimensional polytrope.
It turns out that both options for a sample do not constitute a minimal set of special points to
recover \(C\) from \(\estimateC\), as the following example shows.

\begin{example}\label{ex:dual-covering-4}
    Let \(G = \kappa_4\) and suppose that \(C\) is generic.
    There are two possible cases for the combinatorial structure of \(\wdp(C)\) corresponding to 
    the two central regular triangulations of \(\fundamentalpolytope\), which are shown in \Cref{fig:dual-covering-4}. 

    In both cases, \(\wdp(C)\) has six facets and five pseudovertices, which can be read off
    \Cref{fig:facets-vs-pseudovertices-vs-actual-minimal-sample-size} for \(d=4\) but can also be seen
    from \Cref{fig:dual-covering-4} since \(\fundamentalpolytope\) has six vertices (apart from the origin)
    and both triangulations have five maximal cells.

    Yet for both triangulations the vertex cover given by all maximal cells is not a minimal set
    cover, and a minimal set cover is shown in \Cref{fig:dual-covering-4}.
    One can also see that the minimal size depends on the subdivision and not just the 
    subdivided point configuration.
\end{example}
\begin{figure}[t]
    \centering
    \begin{subfigure}[c]{0.45\textwidth}
        \centering
        \begin{tikzpicture}[
    roundnode/.style={circle, draw=black!85, fill=black!85, thin, scale=0.5},
    pseudonode/.style={circle, draw=black!85, fill=white, scale=0.5},
    sample/.style={draw=blue!85, fill=blue!85, thin},
    scale = 1,
    color = {lightgray}
]

  \coordinate (v0_unnamed__1) at (0, 1);
  \coordinate (v1_unnamed__1) at (0, 0);
  \coordinate (v2_unnamed__1) at (-1, 1);

  \definecolor{vertexcolor_unnamed__1}{rgb}{ 1 0 0 }

  \tikzstyle{vertexstyle_unnamed__1_0} = []
  \tikzstyle{vertexstyle_unnamed__1_1} = []
  \tikzstyle{vertexstyle_unnamed__1_2} = []

  \definecolor{edgecolor_unnamed__1}{rgb}{ 0 0 0 }
  \tikzstyle{facetstyle_unnamed__1} = [fill=black, fill opacity=0.15, draw=edgecolor_unnamed__1, line width=.5 pt, line cap=round, line join=round]

  \draw[facetstyle_unnamed__1] (v1_unnamed__1) -- (v0_unnamed__1) -- (v2_unnamed__1) -- (v1_unnamed__1) -- cycle;

  \foreach \i in {0,1,2} {
    \node at (v\i_unnamed__1) [vertexstyle_unnamed__1_\i] {};
  }

  \coordinate (v0_unnamed__2) at (0, 0);
  \coordinate (v1_unnamed__2) at (-1, 1);
  \coordinate (v2_unnamed__2) at (-1, 0);

  \definecolor{vertexcolor_unnamed__2}{rgb}{ 1 0 0 }

  \tikzstyle{vertexstyle_unnamed__2_0} = []
  \tikzstyle{vertexstyle_unnamed__2_1} = [circle, scale=0.4, fill=orange!85,label={[text=black, above left, align=right]:$\scriptstyle e_{32}$},]
  \tikzstyle{vertexstyle_unnamed__2_2} = []
  
  \definecolor{edgecolor_unnamed__2}{rgb}{ 0 0 0 }
  \tikzstyle{facetstyle_unnamed__2} = [fill=black, fill opacity=0.15, draw=edgecolor_unnamed__2, line width=.5 pt, line cap=round, line join=round]

  \draw[facetstyle_unnamed__2] (v2_unnamed__2) -- (v0_unnamed__2) -- (v1_unnamed__2) -- (v2_unnamed__2) -- cycle;

  \foreach \i in {0,1,2} {
    \node at (v\i_unnamed__2) [vertexstyle_unnamed__2_\i] {};
  }

  \coordinate (v0_unnamed__3) at (0, 0);
  \coordinate (v1_unnamed__3) at (-1, 0);
  \coordinate (v2_unnamed__3) at (0, -1);

  \definecolor{vertexcolor_unnamed__3}{rgb}{ 1 0 0 }

  \tikzstyle{vertexstyle_unnamed__3_0} = []
  \tikzstyle{vertexstyle_unnamed__3_1} = [circle, scale=0.4, fill=orange!85,label={[text=black, above left, align=left]:$\scriptstyle e_{42}$},]
  \tikzstyle{vertexstyle_unnamed__3_2} = []

  \definecolor{edgecolor_unnamed__3}{rgb}{ 0 0 0 }
  \tikzstyle{facetstyle_unnamed__3} = [fill=black, fill opacity=0.15, draw=edgecolor_unnamed__3, line width=.5 pt, line cap=round, line join=round]

  \draw[facetstyle_unnamed__3] (v2_unnamed__3) -- (v1_unnamed__3) -- (v0_unnamed__3) -- (v2_unnamed__3) -- cycle;
  \draw[facetstyle_unnamed__3,orange!85,line width=1pt] (v1_unnamed__2) -- (v1_unnamed__3);

  \foreach \i in {0,1,2} {
    \node at (v\i_unnamed__3) [vertexstyle_unnamed__3_\i] {};
  }

  \coordinate (v0_unnamed__5) at (1, 0);
  \coordinate (v1_unnamed__5) at (0, 0);
  \coordinate (v2_unnamed__5) at (0, -1);

  \definecolor{vertexcolor_unnamed__5}{rgb}{ 1 0 0 }

  \tikzstyle{vertexstyle_unnamed__5_0} = [circle, scale=0.4, fill=blue!85,label={[text=black, above right, align=left]:$\scriptstyle e_{21}$},]
  \tikzstyle{vertexstyle_unnamed__5_1} = [circle, scale=0.4, fill=blue!85,label={[text=black, below right, align=left]:$\scriptstyle e_{41}$},]
  \tikzstyle{vertexstyle_unnamed__5_2} = [circle, scale=0.4, fill=magenta!85,label={[text=black, below right, align=left]:$\scriptstyle e_{43}$},]

  \definecolor{edgecolor_unnamed__5}{rgb}{ 0 0 0 }
  \tikzstyle{facetstyle_unnamed__5} = [fill=black, fill opacity=0.15, draw=edgecolor_unnamed__5, line width=.5 pt, line cap=round, line join=round]

  \draw[facetstyle_unnamed__5] (v2_unnamed__5) -- (v0_unnamed__5) -- (v1_unnamed__5) -- (v2_unnamed__5) -- cycle;

  \foreach \i in {0,1,2} {
    \node at (v\i_unnamed__5) [vertexstyle_unnamed__5_\i] {};
  }

  \coordinate (v0_unnamed__4) at (1, 0);
  \coordinate (v1_unnamed__4) at (0, 1);
  \coordinate (v2_unnamed__4) at (0, 0);

  \definecolor{vertexcolor_unnamed__4}{rgb}{ 1 0 0 }

  \tikzstyle{vertexstyle_unnamed__4_0} = []
  \tikzstyle{vertexstyle_unnamed__4_1} = [circle, scale=0.4, fill=blue!85,label={[text=black, above right, align=left]:$\scriptstyle e_{31}$},]
  \tikzstyle{vertexstyle_unnamed__4_2} = []

  \definecolor{edgecolor_unnamed__4}{rgb}{ 0 0 0 }
  \tikzstyle{facetstyle_unnamed__4} = [fill=blue, fill opacity=0.5, draw=blue, line width=1 pt, line cap=round, line join=round]

  \draw[facetstyle_unnamed__4] (v2_unnamed__4) -- (v0_unnamed__4) -- (v1_unnamed__4) -- (v2_unnamed__4) -- cycle;

  \foreach \i in {0,1,2} {
    \node at (v\i_unnamed__4) [vertexstyle_unnamed__4_\i] {};
  }

\end{tikzpicture}
        \caption{$c_{41} + c_{32} < c_{31} + c_{42}$}
        \label{fig:dual-covering-4:first}
    \end{subfigure}
    \begin{subfigure}[c]{0.45\textwidth}
        \centering
        \begin{tikzpicture}[
    roundnode/.style={circle, draw=black!85, fill=black!85, thin, scale=0.5},
    pseudonode/.style={circle, draw=black!85, fill=white, scale=0.5},
    sample/.style={draw=blue!85, fill=blue!85, thin},
    scale = 1,
    color = {lightgray}
]

  \coordinate (v0_unnamed__1) at (0, 0);
  \coordinate (v1_unnamed__1) at (-1, 0);
  \coordinate (v2_unnamed__1) at (0, -1);

  \definecolor{vertexcolor_unnamed__1}{rgb}{ 1 0 0 }

  \tikzstyle{vertexstyle_unnamed__1_0} = []
  \tikzstyle{vertexstyle_unnamed__1_1} = []
  \tikzstyle{vertexstyle_unnamed__1_2} = []
  
  \definecolor{edgecolor_unnamed__1}{rgb}{ 0 0 0 }
  \tikzstyle{facetstyle_unnamed__1} = [fill=black, fill opacity=0.15, draw=edgecolor_unnamed__1, line width=.5 pt, line cap=round, line join=round]

  \draw[facetstyle_unnamed__1] (v2_unnamed__1) -- (v0_unnamed__1) -- (v1_unnamed__1) -- (v2_unnamed__1) -- cycle;

  \foreach \i in {0,1,2} {
    \node at (v\i_unnamed__1) [vertexstyle_unnamed__1_\i] {};
  }

  \coordinate (v0_unnamed__2) at (0, 1);
  \coordinate (v1_unnamed__2) at (0, 0);
  \coordinate (v2_unnamed__2) at (-1, 0);

  \definecolor{vertexcolor_unnamed__2}{rgb}{ 1 0 0 }

  \tikzstyle{vertexstyle_unnamed__2_0} = []
  \tikzstyle{vertexstyle_unnamed__2_1} = []
  \tikzstyle{vertexstyle_unnamed__2_2} = []

  \definecolor{edgecolor_unnamed__2}{rgb}{ 0 0 0 }
  \tikzstyle{facetstyle_unnamed__2} = [fill=black, fill opacity=0.15, draw=edgecolor_unnamed__2, line width=.5 pt, line cap=round, line join=round]

  \draw[facetstyle_unnamed__2] (v1_unnamed__2) -- (v0_unnamed__2) -- (v2_unnamed__2) -- (v1_unnamed__2) -- cycle;

  \foreach \i in {0,1,2} {
    \node at (v\i_unnamed__2) [vertexstyle_unnamed__2_\i] {};
  }

  \coordinate (v0_unnamed__4) at (1, 0);
  \coordinate (v1_unnamed__4) at (0, 1);
  \coordinate (v2_unnamed__4) at (0, 0);

  \definecolor{vertexcolor_unnamed__4}{rgb}{ 1 0 0 }

  \tikzstyle{vertexstyle_unnamed__4_0} = []
  \tikzstyle{vertexstyle_unnamed__4_1} = []
  \tikzstyle{vertexstyle_unnamed__4_2} = []

  \definecolor{edgecolor_unnamed__4}{rgb}{ 0 0 0 }
  \tikzstyle{facetstyle_unnamed__4} = [fill=black, fill opacity=0.15, draw=edgecolor_unnamed__4, line width=.5 pt, line cap=round, line join=round]

  \draw[facetstyle_unnamed__4] (v2_unnamed__4) -- (v0_unnamed__4) -- (v1_unnamed__4) -- (v2_unnamed__4) -- cycle;

  \foreach \i in {0,1,2} {
    \node at (v\i_unnamed__4) [vertexstyle_unnamed__4_\i] {};
  }

  \coordinate (v0_unnamed__5) at (1, 0);
  \coordinate (v1_unnamed__5) at (0, 0);
  \coordinate (v2_unnamed__5) at (0, -1);

  \tikzstyle{vertexstyle_unnamed__5_0} = [circle, scale=0.4, fill=orange!85,label={[text=black, above right, align=left]:$\scriptstyle e_{21}$},]
  \tikzstyle{vertexstyle_unnamed__5_1} = [circle, scale=0.4, fill=orange!85,label={[text=black, below right, align=left]:$\scriptstyle e_{41}$},]
  \tikzstyle{vertexstyle_unnamed__5_2} = [circle, scale=0.4, fill=orange!85,label={[text=black, below right, align=left]:$\scriptstyle e_{43}$},]

  \definecolor{edgecolor_unnamed__5}{rgb}{ 0 0 0 }
  \tikzstyle{facetstyle_unnamed__5} = [fill=orange, fill opacity=0.5, draw=orange, line width=1 pt, line cap=round, line join=round]

  \draw[facetstyle_unnamed__5] (v2_unnamed__5) -- (v0_unnamed__5) -- (v1_unnamed__5) -- (v2_unnamed__5) -- cycle;

  \foreach \i in {0,1,2} {
    \node at (v\i_unnamed__5) [vertexstyle_unnamed__5_\i] {};
  }

  \coordinate (v0_unnamed__3) at (0, 1);
  \coordinate (v1_unnamed__3) at (-1, 1);
  \coordinate (v2_unnamed__3) at (-1, 0);

  \tikzstyle{vertexstyle_unnamed__3_0} = [circle, scale=0.4, fill=blue!85,label={[text=black, above right, align=left]:$\scriptstyle e_{31}$},]
  \tikzstyle{vertexstyle_unnamed__3_1} = [circle, scale=0.4, fill=blue!85,label={[text=black, above left, align=left]:$\scriptstyle e_{32}$},]
  \tikzstyle{vertexstyle_unnamed__3_2} = [circle, scale=0.4, fill=blue!85,label={[text=black, above left, align=left]:$\scriptstyle e_{42}$},]

  \definecolor{facetcolor_unnamed__3}{rgb}{ 0.4666666667 0.9254901961 0.6196078431 }

  \definecolor{edgecolor_unnamed__3}{rgb}{ 0 0 0 }
  \tikzstyle{facetstyle_unnamed__3} = [fill=blue, fill opacity=0.5, draw=blue, line width=1 pt, line cap=round, line join=round]

  \draw[facetstyle_unnamed__3] (v2_unnamed__3) -- (v0_unnamed__3) -- (v1_unnamed__3) -- (v2_unnamed__3) -- cycle;

  \foreach \i in {0,1,2} {
    \node at (v\i_unnamed__3) [vertexstyle_unnamed__3_\i] {};
  }

\end{tikzpicture}
        \caption{$c_{31} + c_{42} < c_{41} + c_{32}$}
    \end{subfigure}
    \caption{Two central regular triangulations of \(\fundamentalpolytope[\kappa_4]\) from \Cref{ex:dual-covering-4} 
        with a minimal set cover.}
    \label{fig:dual-covering-4}
\end{figure}

This example also shows that the minimum sample size for the estimator \(\widetilde{C}\) depends on the combinatorial type
of \(C\). Thus, for a subdivision of \(\fundamentalpolytope\) denote by \(c(\Sigma)\) the size of a minimum set cover of 
\(\Sigma\). In analogy to set covers for graphs we call this the \emph{set cover number} of \(\Sigma\). 
Then, we define the \emph{minimum sample size} \(c(\thedag)\) for MLBNs supported on \(\thedag\) as
the maximum of the set cover numbers for triangulations of \(\fundamentalpolytope\), \[
    c(\thedag) \coloneqq
    \max\{\,
        c(\Sigma) \mid \Sigma \text{ is a regular central subdivision of }\fundamentalpolytope
    \,\}.
\]

For our computations, we solely considered regular central triangulations of \(\fundamentalpolytope[\kappa_d]\).
This is the maximal case as both the set cover number of a subdivision and the minimum sample size exhibit 
a certain monotonicity in the following sense.

\begin{lemma}\label{lem:monotonicity-set-cover-number}
    Let \(C\in\TT^{d\times d}\) be supported on \(\thedag\) and \(\Sigma\) be the regular subdivision of \(\fundamentalpolytope\)
    induced by \(C\). If \(\Sigma'\) is a subdivision of \(\fundamentalpolytope\) refining \(\Sigma\), then 
    \(c(\Sigma) \leq c(\Sigma')\).

    \begin{proof}
        Suppose \(\widetilde\Sigma'\) is a minimal set cover for \(\Sigma'\). From this, we get a set cover 
        \(\widetilde\Sigma\) for \(\Sigma\) by replacing every cell \(\sigma'\in\widetilde\Sigma'\)
        by the cell \(\sigma\in\Sigma\) which is refined by \(\sigma'\).
        Then, this set cover \(\widetilde\Sigma\) is not necessarily minimal, which gives \(c(\Sigma)\leq c(\Sigma')\).
    \end{proof}
\end{lemma}

\begin{remark}\label{ex:coarsening}
  We can see from \Cref{ex:dual-covering-4} that the bound provided by \Cref{lem:monotonicity-set-cover-number} is not tight
  in general. Denote by \(\Sigma_A\) and \(\Sigma_B\) the two central regular triangulations of \(\fundamentalpolytope[\kappa_4]\)
  that are shown in \Cref{fig:dual-covering-4} and let \(\Sigma'\) be the common coarsening of \(\Sigma_A\) and \(\Sigma_B\).
  The is the subdivision without a diagonal in the upper left square (see \Cref{fig:coarsened-covering}).

  The set cover of \(\Sigma_A\) from \Cref{fig:dual-covering-4:first} translates directly to a set cover of \(\Sigma'\)
  which means that \(c(\Sigma')\leq 3\). But following the proof of \Cref{lem:monotonicity-set-cover-number}, the set cover of \(\Sigma_B\)
  translates to a set cover of \(\Sigma'\) by replacing \[
    \sigma' = \{\,
      e_{13},e_{23},e_{24}
    \,\}\text{ with }
    \sigma = \{\,
      e_{13},e_{14},e_{23},e_{24}
    \,\}.
  \] This gives the upper bound \(c(\Sigma')\leq 2\) which turns out to be sharp in this case.
\end{remark}
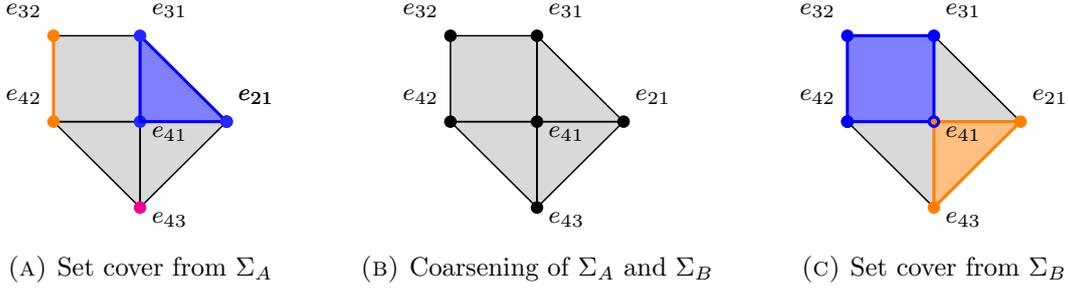
\begin{figure}[tb]
  \centering
  \begin{subfigure}[c]{0.32\textwidth}
    \centering
    \begin{tikzpicture}[
    roundnode/.style={circle, draw=black!85, fill=black!85, thin, scale=0.5},
    pseudonode/.style={circle, draw=black!85, fill=white, scale=0.5},
    sample/.style={draw=blue!85, fill=blue!85, thin},
    scale = 1,
    color = {lightgray}
]

  \coordinate (v0_unnamed__1) at (0, 1);
  \coordinate (v1_unnamed__1) at (0, 0);
  \coordinate (v2_unnamed__1) at (-1, 1);

  \definecolor{vertexcolor_unnamed__1}{rgb}{ 1 0 0 }

  \tikzstyle{vertexstyle_unnamed__1_0} = []
  \tikzstyle{vertexstyle_unnamed__1_1} = []
  \tikzstyle{vertexstyle_unnamed__1_2} = []

  \definecolor{edgecolor_unnamed__1}{rgb}{ 0 0 0 }
  \tikzstyle{facetstyle_unnamed__1} = [fill=black, fill opacity=0.15, draw=edgecolor_unnamed__1, line width=.5 pt, line cap=round, line join=round]


  \foreach \i in {0,1,2} {
    \node at (v\i_unnamed__1) [vertexstyle_unnamed__1_\i] {};
  }

  \coordinate (v0_unnamed__2) at (0, 0);
  \coordinate (v1_unnamed__2) at (-1, 1);
  \coordinate (v2_unnamed__2) at (-1, 0);

  \definecolor{vertexcolor_unnamed__2}{rgb}{ 1 0 0 }

  \tikzstyle{vertexstyle_unnamed__2_0} = []
  \tikzstyle{vertexstyle_unnamed__2_1} = [circle, scale=0.4, fill=orange,label={[text=black, above left, align=right]:$\scriptstyle e_{32}$},]
  \tikzstyle{vertexstyle_unnamed__2_2} = []
  
  \definecolor{edgecolor_unnamed__2}{rgb}{ 0 0 0 }
  \tikzstyle{facetstyle_unnamed__2} = [fill=black, fill opacity=0.15, draw=edgecolor_unnamed__2, line width=.5 pt, line cap=round, line join=round]

  \draw[facetstyle_unnamed__2] (v2_unnamed__2)  -- (v0_unnamed__2) -- (v0_unnamed__1) -- (v1_unnamed__2) -- (v2_unnamed__2) -- cycle;

  \draw[draw=orange,line width=1pt] (v1_unnamed__2) -- (v2_unnamed__2);

  \foreach \i in {0,1,2} {
    \node at (v\i_unnamed__2) [vertexstyle_unnamed__2_\i] {};
  }

  \coordinate (v0_unnamed__3) at (0, 0);
  \coordinate (v1_unnamed__3) at (-1, 0);
  \coordinate (v2_unnamed__3) at (0, -1);

  \definecolor{vertexcolor_unnamed__3}{rgb}{ 1 0 0 }

  \tikzstyle{vertexstyle_unnamed__3_0} = []
  \tikzstyle{vertexstyle_unnamed__3_1} = [circle, scale=0.4, fill=orange,label={[text=black, above left, align=left]:$\scriptstyle e_{42}$},]
  \tikzstyle{vertexstyle_unnamed__3_2} = []

  \definecolor{edgecolor_unnamed__3}{rgb}{ 0 0 0 }
  \tikzstyle{facetstyle_unnamed__3} = [fill=black, fill opacity=0.15, draw=edgecolor_unnamed__3, line width=.5 pt, line cap=round, line join=round]

  \draw[facetstyle_unnamed__3] (v2_unnamed__3) -- (v1_unnamed__3) -- (v0_unnamed__3) -- (v2_unnamed__3) -- cycle;

  \foreach \i in {0,1,2} {
    \node at (v\i_unnamed__3) [vertexstyle_unnamed__3_\i] {};
  }

  \coordinate (v0_unnamed__5) at (1, 0);
  \coordinate (v1_unnamed__5) at (0, 0);
  \coordinate (v2_unnamed__5) at (0, -1);

  \definecolor{vertexcolor_unnamed__5}{rgb}{ 1 0 0 }

  \tikzstyle{vertexstyle_unnamed__5_0} = [circle, scale=0.4, fill=black,label={[text=black, above right, align=left]:$\scriptstyle e_{21}$},]
  \tikzstyle{vertexstyle_unnamed__5_1} = [circle, scale=0.4, fill=blue!85,label={[text=black, below right, align=left]:$\scriptstyle e_{41}$},]
  \tikzstyle{vertexstyle_unnamed__5_2} = [circle, scale=0.4, fill=magenta,label={[text=black, below right, align=left]:$\scriptstyle e_{43}$},]

  \definecolor{edgecolor_unnamed__5}{rgb}{ 0 0 0 }
  \tikzstyle{facetstyle_unnamed__5} = [fill=black, fill opacity=0.15, draw=edgecolor_unnamed__5, line width=.5 pt, line cap=round, line join=round]

  \draw[facetstyle_unnamed__5] (v2_unnamed__5) -- (v0_unnamed__5) -- (v1_unnamed__5) -- (v2_unnamed__5) -- cycle;

  \foreach \i in {0,1,2} {
    \node at (v\i_unnamed__5) [vertexstyle_unnamed__5_\i] {};
  }

  \coordinate (v0_unnamed__4) at (1, 0);
  \coordinate (v1_unnamed__4) at (0, 1);
  \coordinate (v2_unnamed__4) at (0, 0);

  \definecolor{vertexcolor_unnamed__4}{rgb}{ 1 0 0 }

  \tikzstyle{vertexstyle_unnamed__4_0} = [circle, scale=0.4, fill=blue!85,label={[text=black, above right, align=left]:$\scriptstyle e_{21}$},]
  \tikzstyle{vertexstyle_unnamed__4_1} = [circle, scale=0.4, fill=blue!85,label={[text=black, above right, align=left]:$\scriptstyle e_{31}$},]
  \tikzstyle{vertexstyle_unnamed__4_2} = []

  \definecolor{edgecolor_unnamed__4}{rgb}{ 0 0 0 }
  \tikzstyle{facetstyle_unnamed__4} = [fill=blue, fill opacity=0.5, draw=blue, line width=1pt, line cap=round, line join=round]

  \draw[facetstyle_unnamed__4] (v2_unnamed__4) -- (v0_unnamed__4) -- (v1_unnamed__4) -- (v2_unnamed__4) -- cycle;

  \foreach \i in {0,1,2} {
    \node at (v\i_unnamed__4) [vertexstyle_unnamed__4_\i] {};
  }

\end{tikzpicture}
    \caption{Set cover from \(\subdivision_A\)}
  \end{subfigure}
  \begin{subfigure}[c]{0.32\textwidth}
    \centering
    \begin{tikzpicture}[
    roundnode/.style={circle, draw=black!85, fill=black!85, thin, scale=0.5},
    pseudonode/.style={circle, draw=black!85, fill=white, scale=0.5},
    sample/.style={draw=blue!85, fill=blue!85, thin},
    scale = 1,
    color = {lightgray}
]

  \coordinate (v0_unnamed__1) at (0, 1);
  \coordinate (v1_unnamed__1) at (0, 0);
  \coordinate (v2_unnamed__1) at (-1, 1);

  \definecolor{vertexcolor_unnamed__1}{rgb}{ 1 0 0 }

  \tikzstyle{vertexstyle_unnamed__1_0} = []
  \tikzstyle{vertexstyle_unnamed__1_1} = []
  \tikzstyle{vertexstyle_unnamed__1_2} = []

  \definecolor{edgecolor_unnamed__1}{rgb}{ 0 0 0 }
  \tikzstyle{facetstyle_unnamed__1} = [fill=black, fill opacity=0.15, draw=edgecolor_unnamed__1, line width=.5 pt, line cap=round, line join=round]


  \foreach \i in {0,1,2} {
    \node at (v\i_unnamed__1) [vertexstyle_unnamed__1_\i] {};
  }

  \coordinate (v0_unnamed__2) at (0, 0);
  \coordinate (v1_unnamed__2) at (-1, 1);
  \coordinate (v2_unnamed__2) at (-1, 0);

  \definecolor{vertexcolor_unnamed__2}{rgb}{ 1 0 0 }

  \tikzstyle{vertexstyle_unnamed__2_0} = []
  \tikzstyle{vertexstyle_unnamed__2_1} = [circle, scale=0.4, fill=black,label={[text=black, above left, align=right]:$\scriptstyle e_{32}$},]
  \tikzstyle{vertexstyle_unnamed__2_2} = []
  
  \definecolor{edgecolor_unnamed__2}{rgb}{ 0 0 0 }
  \tikzstyle{facetstyle_unnamed__2} = [fill=black, fill opacity=0.15, draw=edgecolor_unnamed__2, line width=.5 pt, line cap=round, line join=round]

  \draw[facetstyle_unnamed__2] (v2_unnamed__2)  -- (v0_unnamed__2) -- (v0_unnamed__1) -- (v1_unnamed__2) -- (v2_unnamed__2) -- cycle;

  \foreach \i in {0,1,2} {
    \node at (v\i_unnamed__2) [vertexstyle_unnamed__2_\i] {};
  }

  \coordinate (v0_unnamed__3) at (0, 0);
  \coordinate (v1_unnamed__3) at (-1, 0);
  \coordinate (v2_unnamed__3) at (0, -1);

  \definecolor{vertexcolor_unnamed__3}{rgb}{ 1 0 0 }

  \tikzstyle{vertexstyle_unnamed__3_0} = []
  \tikzstyle{vertexstyle_unnamed__3_1} = [circle, scale=0.4, fill=black,label={[text=black, above left, align=left]:$\scriptstyle e_{42}$},]
  \tikzstyle{vertexstyle_unnamed__3_2} = []

  \definecolor{edgecolor_unnamed__3}{rgb}{ 0 0 0 }
  \tikzstyle{facetstyle_unnamed__3} = [fill=black, fill opacity=0.15, draw=edgecolor_unnamed__3, line width=.5 pt, line cap=round, line join=round]

  \draw[facetstyle_unnamed__3] (v2_unnamed__3) -- (v1_unnamed__3) -- (v0_unnamed__3) -- (v2_unnamed__3) -- cycle;

  \foreach \i in {0,1,2} {
    \node at (v\i_unnamed__3) [vertexstyle_unnamed__3_\i] {};
  }

  \coordinate (v0_unnamed__5) at (1, 0);
  \coordinate (v1_unnamed__5) at (0, 0);
  \coordinate (v2_unnamed__5) at (0, -1);

  \definecolor{vertexcolor_unnamed__5}{rgb}{ 1 0 0 }

  \tikzstyle{vertexstyle_unnamed__5_0} = [circle, scale=0.4, fill=black,label={[text=black, above right, align=left]:$\scriptstyle e_{21}$},]
  \tikzstyle{vertexstyle_unnamed__5_1} = [circle, scale=0.4, fill=black,label={[text=black, below right, align=left]:$\scriptstyle e_{41}$},]
  \tikzstyle{vertexstyle_unnamed__5_2} = [circle, scale=0.4, fill=black,label={[text=black, below right, align=left]:$\scriptstyle e_{43}$},]

  \definecolor{edgecolor_unnamed__5}{rgb}{ 0 0 0 }
  \tikzstyle{facetstyle_unnamed__5} = [fill=black, fill opacity=0.15, draw=edgecolor_unnamed__5, line width=.5 pt, line cap=round, line join=round]

  \draw[facetstyle_unnamed__5] (v2_unnamed__5) -- (v0_unnamed__5) -- (v1_unnamed__5) -- (v2_unnamed__5) -- cycle;

  \foreach \i in {0,1,2} {
    \node at (v\i_unnamed__5) [vertexstyle_unnamed__5_\i] {};
  }

  \coordinate (v0_unnamed__4) at (1, 0);
  \coordinate (v1_unnamed__4) at (0, 1);
  \coordinate (v2_unnamed__4) at (0, 0);

  \definecolor{vertexcolor_unnamed__4}{rgb}{ 1 0 0 }

  \tikzstyle{vertexstyle_unnamed__4_0} = []
  \tikzstyle{vertexstyle_unnamed__4_1} = [circle, scale=0.4, fill=black,label={[text=black, above right, align=left]:$\scriptstyle e_{31}$},]
  \tikzstyle{vertexstyle_unnamed__4_2} = []

  \definecolor{edgecolor_unnamed__4}{rgb}{ 0 0 0 }
  \tikzstyle{facetstyle_unnamed__4} = [fill=black, fill opacity=0.15, draw=black, line width=.5 pt, line cap=round, line join=round]

  \draw[facetstyle_unnamed__4] (v2_unnamed__4) -- (v0_unnamed__4) -- (v1_unnamed__4) -- (v2_unnamed__4) -- cycle;

  \foreach \i in {0,1,2} {
    \node at (v\i_unnamed__4) [vertexstyle_unnamed__4_\i] {};
  }

\end{tikzpicture}
    \caption{Coarsening of \(\subdivision_A\) and \(\subdivision_B\)}
  \end{subfigure}
  \begin{subfigure}[c]{0.32\textwidth}
    \centering
    \begin{tikzpicture}[
    roundnode/.style={circle, draw=black!85, fill=black!85, thin, scale=0.5},
    pseudonode/.style={circle, draw=black!85, fill=white, scale=0.5},
    sample/.style={draw=blue!85, fill=blue!85, thin},
    scale = 1,
    color = {lightgray}
]

  \coordinate (v0_unnamed__1) at (0, 1);
  \coordinate (v1_unnamed__1) at (0, 0);
  \coordinate (v2_unnamed__1) at (-1, 1);

  \definecolor{vertexcolor_unnamed__1}{rgb}{ 1 0 0 }

  \tikzstyle{vertexstyle_unnamed__1_0} = []
  \tikzstyle{vertexstyle_unnamed__1_1} = []
  \tikzstyle{vertexstyle_unnamed__1_2} = []

  \definecolor{edgecolor_unnamed__1}{rgb}{ 0 0 0 }
  \tikzstyle{facetstyle_unnamed__1} = [fill=black, fill opacity=0.15, draw=edgecolor_unnamed__1, line width=.5 pt, line cap=round, line join=round]


  \foreach \i in {0,1,2} {
    \node at (v\i_unnamed__1) [vertexstyle_unnamed__1_\i] {};
  }

  \coordinate (v0_unnamed__3) at (0, 0);
  \coordinate (v1_unnamed__3) at (-1, 0);
  \coordinate (v2_unnamed__3) at (0, -1);

  \definecolor{vertexcolor_unnamed__3}{rgb}{ 1 0 0 }

  \tikzstyle{vertexstyle_unnamed__3_0} = []
  \tikzstyle{vertexstyle_unnamed__3_1} = [circle, scale=0.4, fill=black,label={[text=black, above left, align=left]:$\scriptstyle e_{42}$},]
  \tikzstyle{vertexstyle_unnamed__3_2} = []

  \definecolor{edgecolor_unnamed__3}{rgb}{ 0 0 0 }
  \tikzstyle{facetstyle_unnamed__3} = [fill=black, fill opacity=0.15, draw=edgecolor_unnamed__3, line width=.5 pt, line cap=round, line join=round]

  \draw[facetstyle_unnamed__3] (v2_unnamed__3) -- (v1_unnamed__3) -- (v0_unnamed__3) -- (v2_unnamed__3) -- cycle;

  \foreach \i in {0,1,2} {
    \node at (v\i_unnamed__3) [vertexstyle_unnamed__3_\i] {};
  }

  \coordinate (v0_unnamed__4) at (1, 0);
  \coordinate (v1_unnamed__4) at (0, 1);
  \coordinate (v2_unnamed__4) at (0, 0);

  \definecolor{vertexcolor_unnamed__4}{rgb}{ 1 0 0 }

  \tikzstyle{vertexstyle_unnamed__4_0} = []
  \tikzstyle{vertexstyle_unnamed__4_1} = [circle, scale=0.4, fill=blue,label={[text=black, above right, align=left]:$\scriptstyle e_{31}$},]
  \tikzstyle{vertexstyle_unnamed__4_2} = []

  \definecolor{edgecolor_unnamed__4}{rgb}{ 0 0 0 }
  \tikzstyle{facetstyle_unnamed__4} = [fill=black, fill opacity=0.15, draw=black, line width=.5 pt, line cap=round, line join=round]

  \draw[facetstyle_unnamed__4] (v2_unnamed__4) -- (v0_unnamed__4) -- (v1_unnamed__4) -- (v2_unnamed__4) -- cycle;

  \foreach \i in {0,1,2} {
    \node at (v\i_unnamed__4) [vertexstyle_unnamed__4_\i] {};
  }

  \coordinate (v0_unnamed__2) at (0, 0);
  \coordinate (v1_unnamed__2) at (-1, 1);
  \coordinate (v2_unnamed__2) at (-1, 0);

  \definecolor{vertexcolor_unnamed__2}{rgb}{ 1 0 0 }

  \tikzstyle{vertexstyle_unnamed__2_0} = [circle, scale=.2,fill=orange]]
  \tikzstyle{vertexstyle_unnamed__2_1} = [circle, scale=0.4, fill=blue,label={[text=black, above left, align=right]:$\scriptstyle e_{32}$},]
  \tikzstyle{vertexstyle_unnamed__2_2} = [circle, scale=.4, fill=blue]
  
  \definecolor{edgecolor_unnamed__2}{rgb}{ 0 0 0 }
  \tikzstyle{facetstyle_unnamed__2} = [fill=blue, fill opacity=0.5, draw=blue, line width=1pt, line cap=round, line join=round]

  \draw[facetstyle_unnamed__2] (v2_unnamed__2)  -- (v0_unnamed__2) -- (v0_unnamed__1) -- (v1_unnamed__2) -- (v2_unnamed__2) -- cycle;

  \foreach \i in {0,1,2} {
    \node at (v\i_unnamed__2) [vertexstyle_unnamed__2_\i] {};
  }

  \coordinate (v0_unnamed__5) at (1, 0);
  \coordinate (v1_unnamed__5) at (0, 0);
  \coordinate (v2_unnamed__5) at (0, -1);

  \definecolor{vertexcolor_unnamed__5}{rgb}{ 1 0 0 }

  \tikzstyle{vertexstyle_unnamed__5_0} = [circle, scale=0.4, fill=orange,label={[text=black, above right, align=left]:$\scriptstyle e_{21}$},]
  \tikzstyle{vertexstyle_unnamed__5_1} = [circle, scale=0.4, fill=blue,label={[text=black, below right, align=left]:$\scriptstyle e_{41}$},]
  \tikzstyle{vertexstyle_unnamed__5_2} = [circle, scale=0.4, fill=orange,label={[text=black, below right, align=left]:$\scriptstyle e_{43}$},]

  \definecolor{edgecolor_unnamed__5}{rgb}{ 0 0 0 }
  \tikzstyle{facetstyle_unnamed__5} = [fill=orange, fill opacity=0.5, draw=orange, line width=1pt, line cap=round, line join=round]

  \draw[facetstyle_unnamed__5] (v2_unnamed__5) -- (v0_unnamed__5) -- (v1_unnamed__5) -- (v2_unnamed__5) -- cycle;

  \foreach \i in {0,1,2} {
    \node at (v\i_unnamed__5) [vertexstyle_unnamed__5_\i] {};
  }
  \node at (v0_unnamed__2) [vertexstyle_unnamed__2_0] {};
\end{tikzpicture}
    \caption{Set cover from \(\subdivision_B\)}
  \end{subfigure}
  \caption{Coarse subdivision from \Cref{ex:coarsening} together with set covers obtained from the triangulations 
  from \Cref{ex:dual-covering-4} by applying \Cref{lem:monotonicity-set-cover-number}.}
  \label{fig:coarsened-covering}
\end{figure}

\begin{lemma}
    Let \(\subdag\subseteq\thedag\) be DAGs. Then, \(c(\subdag)\leq c(\thedag)\).

    \begin{proof}
      Let \(\subdivision'\) be a central regular triangulation of \(\fundamentalpolytope[\subdag]\) realizing the minimum sample size for 
      \(\subdag\), meaning \(c(\subdivision') = c(\subdag)\). Assume first that \(\subdag\) is equal to \(\thedag\) with 
      an edge \(j\to i\) removed.

      We can extend \(\subdivision'\) into a subdivision \(\subdivision\) for \(\fundamentalpolytope\) by replacing every
      cell \(\sigma'\in\subdivision'\) \emph{visible} from \(e_{ji}\) with \(\sigma=\conv(\sigma'\cup\{\,e_{ji}\,\})\). 
      By visible we mean that the line segment connecting \(e_{ji}\) with the barycenter of \(\sigma'\) intersects \(\Sigma'\) 
      only in \(\sigma'\), see \Cref{fig:visibility}. In particular, a minimum set cover for \(\subdivision'\)
      stays a minimum set cover for \(\subdivision\) by replacing any \(\sigma'\) as above with \(\sigma\).

      This subdivision \(\subdivision\) is not a triangulation of \(\fundamentalpolytope\), but still central and regular
      and as such is refined by a central regular triangulation \(\subdivision''\) of \(\fundamentalpolytope\).
      \Cref{lem:monotonicity-set-cover-number} implies that \(c(\subdivision)\leq c(\subdivision'')\).
      Summarizing the above, we get \[
        c(\subdag) = c(\subdivision') = c(\subdivision) \leq c(\subdivision'') \leq c(\thedag).
      \] We can now generalize this to arbitrary pairs \(\subdag\subseteq\thedag\) by using the above argument inductively
      for each edge in \(\thedag\setminus\subdag\).
      \end{proof}
\end{lemma}

\begin{figure}[b]
  \centering
  \hfill
  \begin{tikzpicture}[
    roundnode/.style={circle, draw=black!85, fill=black!85, thin, scale=0.4},
    barycenter/.style={regular polygon, regular polygon sides=3, fill=black!85, draw=black!85, scale=0.15},
    cross/.style={cross out, fill=black!85, draw=black!85, scale=0.3},
    scale = 1,
    color = {lightgray}
]
  \coordinate (14) at (0, 0);
  \coordinate (13) at (0, 1);
  \coordinate (12) at (1, 0);
  \coordinate (23) at (-1,1);
  \coordinate (24) at (-1,0);
  \coordinate (34) at (0,-1);

  \draw[fill=black, fill opacity=.15, draw=black, line width=.5pt, line cap=round, line join=round] (14) -- (13) -- (23) -- (24) -- cycle;
  \draw[black, line width=.5pt] (13) -- (24);

  \node[roundnode,label={[text=black,above right, align=left]:$\scriptstyle e_{21}$}] at (12) {};
  \node[roundnode,label={[text=black,below right, align=left]:$\scriptstyle e_{41}$}] at (14) {};
  \node[roundnode,label={[text=black,above left, align=left]:$\scriptstyle e_{42}$}] at (24) {};
  \node[roundnode,label={[text=black,above left, align=left]:$\scriptstyle e_{32}$}] at (23) {};
  \node[roundnode,label={[text=black,above right, align=left]:$\scriptstyle e_{31}$}] at (13) {};

  \coordinate (b1) at (-.33,.33);
  \coordinate (b2) at (-.66,.66);
  \node[barycenter] at (b1) {};
  \node[barycenter] at (b2) {};

  \draw[black, line width=.5pt, dashed] (12) -- (b1);
  \draw[black, line width=.5pt, dashed] (12) -- node[cross,pos=.6,orange] {} (b2);
\end{tikzpicture}
  \hfill
  \begin{tikzpicture}[
    roundnode/.style={circle, draw=black!85, fill=black!85, thin, scale=0.4},
    barycenter/.style={regular polygon, regular polygon sides=3, fill=black!85, draw=black!85, scale=0.15},
    cross/.style={cross out, fill=black!85, draw=black!85, scale=0.3},
    scale = 1,
    color = {lightgray}
]
  \coordinate (14) at (0, 0);
  \coordinate (13) at (0, 1);
  \coordinate (12) at (1, 0);
  \coordinate (23) at (-1,1);
  \coordinate (24) at (-1,0);
  \coordinate (34) at (0,-1);

  \draw[fill=black, fill opacity=.15, draw=black, line width=.5pt, line cap=round, line join=round] (14) -- (24) -- (13) -- (12) -- cycle;
  \draw[fill=black, fill opacity=.15, draw=black, line width=.5pt, line cap=round, line join=round] (13) -- (23) -- (24) -- cycle;

  \node[roundnode,label={[text=black,above right, align=left]:$\scriptstyle e_{21}$}] at (12) {};
  \node[roundnode,label={[text=black,below right, align=left]:$\scriptstyle e_{41}$}] at (14) {};
  \node[roundnode,label={[text=black,above left, align=left]:$\scriptstyle e_{42}$}] at (24) {};
  \node[roundnode,label={[text=black,above left, align=left]:$\scriptstyle e_{32}$}] at (23) {};
  \node[roundnode,label={[text=black,above right, align=left]:$\scriptstyle e_{31}$}] at (13) {};
\end{tikzpicture}
  \hfill
  \caption{Extending a subdivision \(\subdivision'\) for \(\fundamentalpolytope[\subdag]\) to a subdivision
  \(\subdivision\) for \(\fundamentalpolytope\) by extending the cells visible from a new point \(e_{ji}\).}
  \label{fig:visibility}
\end{figure}

For moderate \(d\) we have computationally investigated the minimum set covers of triangulations for 
\(\fundamentalpolytope[\kappa_d]\). We employed a simple randomized greedy algorithm (\Cref{alg:greedy-set-cover})
with \(100\) repetitions to calculate a set cover for each triangulation. 
For each triangulation this yields a minimal observed cardinality of a set cover and 
for each \(d\) the observed minima are shown in \Cref{tab:enumeration}.

For \(d\leq 7\) we were able to iterate through all triangulations of \(\fundamentalpolytope[\kappa_d]\), 
while for \(d\geq 8\) we decided to sample a 5000 triangulations and find the size of minimal set covers 
with above randomized procedure. 

\begin{table}[h]
   \centering
   \caption{Observed set cover numbers across combinatorial types of \(\wdp(C)\) for generic \(C\in\TT^{d\times d}\)
   supported on \(\thedag = \kappa_d\). For \(^{(*)}\), the observed set cover numbers are obtained from 
   a sample of \(5000\) triangulations.}
   \begin{tabular}{lccccccccc}
       \toprule
       $d$ & 1 & 2 & 3 &   4 & 5 &   6 &    7 & $8^{(*)}$ & $9^{(*)}$ \\\midrule
           & 0 & 1 & 2 & 2,3 & 3 & 3,4 &  4,5 & 5,6       & 5,6,7     \\
       \bottomrule
   \end{tabular}
   \label{tab:enumeration}
\end{table}

\begin{algorithm}[t]
\caption{Randomized greedy set cover for triangulations}
\label{alg:greedy-set-cover}
\begin{algorithmic}[1]
\Input triangulation \(\Sigma\) for a point configuration \(P\)
\State \(\widetilde\Sigma\leftarrow\varnothing\)
\While{there exists \(p\in P\) not covered by \(\widetilde\Sigma\)}
    \State Choose \(\sigma\in\Sigma\) with the least covered elements \(p\in P\)
    \State \(\widetilde\Sigma\leftarrow\widetilde\Sigma\cup\{\sigma\}\)
\EndWhile
\State\Return\(\widetilde\Sigma\)
\end{algorithmic}
\end{algorithm}

\section{Structural learning and gaps}\label{sec:structural-learning}
In the regime where we do not know the underlying DAG \(\thedag\) but do know that the 
ordering of the coordinates corresponds to a topological ordering of \(\thedag\), we find ourselves 
in a slight variation of the situation where we assume \(\thedag = \kappa_d\). 

In the worst case we need to estimate as many edges as in the case when \(\thedag = \kappa_d\) is actually known,
but we have to estimate the existence an edge additionally. Thus, we cannot obtain an estimate \(\estimateC\) by simply
calculating \(\widetilde{C}\) and restricting to entries corresponding to known edges.

A simple condition to check whether the maximum of \(X_{ij}\) corresponds to an atom, and thus an edge, is to check
whether \(X_{ij} = \widetilde{c}_{ij}\) multiple times in a given sample \cite{GKL:2021}. Alternatively, there are more
involved scoring methods based on concentration measures for \(X_{ij}\) that have been shown by 
\citeauthorandcite{BKT:2024} to work in the case of MLBNs with underlying DAG being a tree.

To \emph{score} whether an observed maximum of \(X_{ij}\) corresponds to an edge, we want a non-negative function \(s\) 
that given the observations for \(X_{ij}\) produces a small number if there is a high concentration
around the maximum of \(X_{ij}\), indicating an atom and thus the presence of an edge.
The following are some possible choices for \emph{scoring functions}:

\begin{itemize}
    \item The \emph{top-\(k\) gap} that compares the maximum of \(X_{ij}\) with the \(k\)-th highest observation
        (assuming that the observations of \(X_{ij}\) are sorted from largest to smallest),
        \[
            s_{ij}(k) \coloneqq \left(
                X_{ij}^{(1)} - X_{ij}^{(k)}
            \right)^2.
        \]

    \item The empirical \emph{quantile-to-mean gap} that compares an upper quantile with the mean,
        \[
            s_{ij}(\overline{r}) \coloneqq 
                \frac
                    {(\mathbb{E}[X_{ij}] - Q_{X_{ij}}(\overline{r}))^2}
                    {n_{ij}}.
        \]

    \item The empirical \emph{upper quantile gap} that compares two fixed quantile levels,
        \[
            s_{ij}(\overline{r},\underline{r}) \coloneqq 
                \frac
                    {(Q_{X_{ij}}(\overline{r}) - Q_{X_{ij}}(\underline{r}))^2}
                    {n_{ij}}.
        \]
\end{itemize}

If we additionally assume that the ordering of the coordinates is not necessarily known, we additionally have to 
decide whether the maximum of \(X_{ij}\) or the maximum of \(X_{ji}\) (which is the minimum of \(X_{ij}\)) exhibits
an atom and which one scores better. For this, we combine the scoring approach from \citeauthor{BKT:2024}
with thresholding proposed by \citeauthorandcite{BK:2021}.
These considerations are put together in \Cref{alg:estimate-with-ordering} to calculate an estimate \(\estimateC\) 
given a sample \(S\) from a MLBN by calculating \(\widetilde{C}\) for the minimum bounding polytrope of \(S\)
and pruning edges.

\begin{algorithm}[h]
\caption{Estimating \(\estimateC\) for a MLBN with unknown DAG and ordering}
\label{alg:estimate-with-ordering}
\begin{algorithmic}[1]
\Input Sample $S := \{\,p^{(1)}, \ldots , p^{(n)}\,\}$ originating from an MLBN
\Parameter threshold \(t>0\), a scoring function \(s\)
\State \(\estimateC\leftarrow \left(\max_kp^{(k)}_i - p^{(k)}_j\right)_{1\leq i,j\leq d}\)
    \Comment{minimum bounding polytrope for \(S\)}
\State Compute the scores \(s_{ij}\)
\For{$1\leq i\neq j\leq d$}
    \If{$s_{ij},s_{ji}\geq t$}
        \State \(\widehat{c}_{ij},\widehat{c}_{ji} \leftarrow \infty \) 
            \Comment{Neither direction between \(i\) and \(j\) is present}
    \ElsIf{$s_{ij} < s_{ji}$}
        \State \(\widehat{c}_{ji} \leftarrow \infty \) \Comment{\(j\to i\) is present}
    \Else
        \State \(\widehat{c}_{ij} \leftarrow \infty \) \Comment{\(i\to j\) is present}
    \EndIf
\EndFor
\State \Return \(\estimateC\)
\end{algorithmic}
\end{algorithm}

\section{Experiments on data}\label{sec:experiments}
We test \Cref{alg:estimate-with-ordering} in several settings, on simulated data and on two real-world datasets.
The code to reproduce our experiments is available under the following link: \begin{center}
  \url{https://zenodo.org/records/17054748}
\end{center}
To estimate the performance of our method, we use standard performance metrics in causal inference \cite{ZARX:2018}
that have also been used in \cite{BKT:2024} for the evaluation of the methods therein.
We recall the definitions of those in the following.

Denote by \(\mathrm{SHD}(\thedag,\estimate)\) the \emph{structural Hamming distance} between \(\thedag\) and 
the estimate \(\estimate\), which is the minimal number of edge additions, deletions and reversals necessary 
to turn \(\estimate\) into \(\thedag\). Denote by \(E\) and \(\estimateEdges\) the set of edges of \(\thedag\) 
respectively \(\estimate\). Then, \(E\cap\estimateEdges\) is the set of correctly estimated edges while
\(\estimateEdges\setminus E\) contains the edges that have been estimated incorrectly, either by reversal of direction
or not being present in \(\thedag\). Applying appropriate normalizing factors, we obtain the following four 
quantities:
\begin{align*}
    \mathrm{nSHD}(\thedag,\widetilde\thedag)&\coloneq\frac{\mathrm{SHD}(\thedag,\widetilde\thedag)}{\#E+\#\widetilde{E}},
    &\mathrm{FDR}(\thedag,\widetilde\thedag)&\coloneq\frac{\#(\widetilde{E}\setminus E)}{\#\widetilde{E}},\\
    \mathrm{FPR}(\thedag,\widetilde\thedag)&\coloneq\frac{\#(\widetilde{E}\setminus E)}{d(d-1)-\#E},
    &\mathrm{TPR}(\thedag,\widetilde\thedag)&\coloneq\frac{\#(E\cap\widetilde{E})}{\#E}.
\end{align*}
The \emph{false discovery rate} FDR describes the ratio of incorrectly estimated edges among all estimated edges,
the \emph{false positive rate} FPR describes the ratio of incorrectly estimated edges compared to all non-present
edges in \(\thedag\) while the \emph{true positive rate} is the ratio of correctly estimated edges.

\subsection{Simulated data}\label{sec:simulation}
For our simulation study, we have generated data \(S\) from a max-linear Bayesian network using \eqref{eq:MLBN-solution}
assuming \(Z_i\sim D\) i.\,i.\,d.\ for some distribution \(D\). For the parameter matrix \(C\), we draw each weight
\(c_{ij}\) uniformly from a fixed interval \([-\tau,\tau]\) for some \(\tau\geq 0\) and select each edge \(j\to i\in\thedag\)
with fixed probability \(p\in(0,1]\). This way, we obtain a lower-triangular matrices \(C\in\TT^{d\times d}\) as
parameter matrix.

We performed simulations for both the setting of known DAG \(\thedag\) as in \Cref{sec:param-recovery}
and unknown DAG with unknown ordering as in \Cref{sec:structural-learning}. For the latter setting, 
we additionally generated a permutation \(\sigma\in\SymGr_d\) to obtain a random ordering of the nodes of \(\thedag\).
We simulated for graphs of sizes \(d=5,10,30,50,100\) with \(5000\) repetitions with sample size \(N = 1000\) for 
each simulation. For the scoring functions, we have used the top-k gap with \( k = 30 \) and the quantile-to-mean gap
with \(\overline{r} = 0.95\).

We considered the following settings for the distributions of the innovations \(Z_i\) and selection of the model parameters:
\begin{enumerate}
    \item Gaussian i.\,i.\,d.\ innovations and \(G = \kappa_d\) $(p = 1)$,
    \item Gaussian i.\,i.\,d.\ innovations and \(p = \frac12\), 
    \item Fr\'echet i.\,i.\,d.\ innovations and \(G = \kappa_d\) $(p = 1)$, and
    \item Fr\'echet i.\,i.\,d.\ innovations and \(p = \frac12\).
\end{enumerate}
Each setting was simulated with a fixed ordering where \(C\) is lower-triangular 
(\Cref{tab:simulation-study-metrics:fixed-ordering} and \Cref{tab:simulation-study-metrics:fixed-ordering:quantile-to-mean}), 
and with a random ordering of the coordinates (\Cref{tab:simulation-study-metrics:random-ordering} and \Cref{tab:simulation-study-metrics:random-ordering:quantile-to-mean}).

Our results show that \Cref{alg:estimate-with-ordering} performs reasonably well for MLBNs of moderate size, that is
with underlying DAGs on \(d\leq30\) nodes, when using the top-k gap as scoring function. In this case, the performance
is slightly worse for Fr\'echet distributed innovations than for Gaussian distributed ones. Also, the algorithm with 
top-k gap performs similarly for instances with known ordering and for instances with random, unknown ordering.

When using the quantile-to-mean gap as scoring function, the performance of \Cref{alg:estimate-with-ordering} is worse than
using top-k gap. For large graphs ($d=50,100$), the true positive rate is close to 50\% indicating that this variant
performs similar to randomly guessing the edges. Additionally, the performance for Fr\'echet distributed innovations
is worse which could be due to the presence of extreme outliers.
This also supports our observation that when using the quantile-to-mean gap as scoring function the threshold in 
\Cref{alg:estimate-with-ordering} has to be adjusted to the specific scenario.

\subsection{Dietary supplement data}
In this experiment, we consider dietary supplement data of $n = 8327$ independent interviews taken from the 
NHANES report for the period 2015--2016. This data has been originally available from \begin{center}
    \url{https://wwwn.cdc.gov/Nchs/Nhanes/2015-2016/DR1TOT_I.XPT}
\end{center} but at the time of writing this dataset has not been available from its original address and instead
had to be retrieved from the Internet Archive. 

This dataset has been used in \cite{KK:2021,BK:2021} to test the performance of the therein developed estimators
for MLBNs. In both cases the focus was on four of the 168 data components which describe the intake of 
Vitamin A (\texttt{DR1TVARA}), Beta-Carotene (\texttt{DR1TBCAR}), Lutein+Zeaxanthin (\texttt{DR1TLZ}) 
and Alpha-Carotene (\texttt{DR1TACAR}). We abbreviate these components as VA, BC, LZ, AC respectively.
For preprocessing, we only applied a negative logarithmic transformation to these four components.

For \Cref{alg:estimate-with-ordering} together with the top-k gap, we recovered the following DAG.
We additionally give the Kleene star \(\estimateC^\star\) of our estimate for sake of comparability with
\cite{KK:2021,BK:2021}.
The DAG agrees with the one obtained by \cite[Section 6.1]{BK:2021} but is missing the edge \(\text{AC}\to\text{LZ}\)
of the DAG estimated by \cite[Section 7.3]{KK:2021}.
\[
    \begin{tikzcd}[row sep=1.5em]
        & AC \arrow[ld]\arrow[dd] & \\
        BC \arrow[rd] & & LZ \arrow[ll]\arrow[ld] \\
        & VA &
    \end{tikzcd}
    \quad\text{with estimate}\quad
    \estimateC^\star = \begin{blockarray}{ccccc}
        \text{VA} & \text{AC} & \text{BC} & \text{LZ} \\
        \begin{block}{(cccc)c}
          0      & 2.06   & 2.56   & 6.73   & \text{VA} \\
          \infty & 0      & \infty & \infty & \text{AC} \\
          \infty & 0.25   & 0      & 4.17   & \text{BC} \\
          \infty & \infty & \infty & 0      & \text{LZ} \\
        \end{block}
    \end{blockarray}
\]

When using the empirical quantile-to-mean gap as scoring function, we obtained a directed graph with a cycle.
This shows an important caveat about our method since there is no guarantee that the estimated 
graph contains no cycles.
\[
    \begin{tikzcd}[row sep=1.5em]
        & AC \arrow[ld] & \\
        BC \arrow[rd] & & LZ \arrow[ll] \\
        & VA\arrow[ur] &
    \end{tikzcd}
    \quad\text{with estimate}\quad
    \estimateC^\star = \begin{blockarray}{ccccc}
        \text{VA} & \text{AC} & \text{BC} & \text{LZ} \\
        \begin{block}{(cccc)c}
          0      & \infty & 2.56   & \infty & \text{VA} \\
          \infty & 0      & \infty & \infty & \text{AC} \\
          \infty & 0.25   & 0      & 4.17   & \text{BC} \\
          7.81   & \infty & \infty & 0      & \text{LZ} \\
        \end{block}
    \end{blockarray}
\]

\subsection{The upper Danube network}
The Danube network data consist of daily measurements collected at d = 31 gauging stations in the Danube river basin 
over 50 years from 1960 to 2009 by the Bavarian Environmental Agency. 
This region is to prone to serious flooding which provides the extreme events to apply MLBNs to.

This dataset has been a mainstay in benchmarking statistical methods for estimating
MLBNs \cite{EH:2020,BKT:2024}. We have used the version of this dataset included in \cite{R:graphicalExtremes}
which incorporates the preprocessing by \cite{ADE:2015}. This includes restricting the data to the months
June, July and August since the most extreme flooding events occur at this time due to heavy summer rain \cite{BW:2006}.

While fitting using the top-k gap and quantile-to-mean gap as scoring method, we did not recover the true graph
successfully. While our method recovered some edges also present in the true network, there is also a large amount
of edges in the estimated DAG either in the wrong direction or not present in the true network.
The results for both attempts are given in \Cref{tab:danube}.

\begin{figure}
    \centering
    \includegraphics[width=0.6\linewidth]{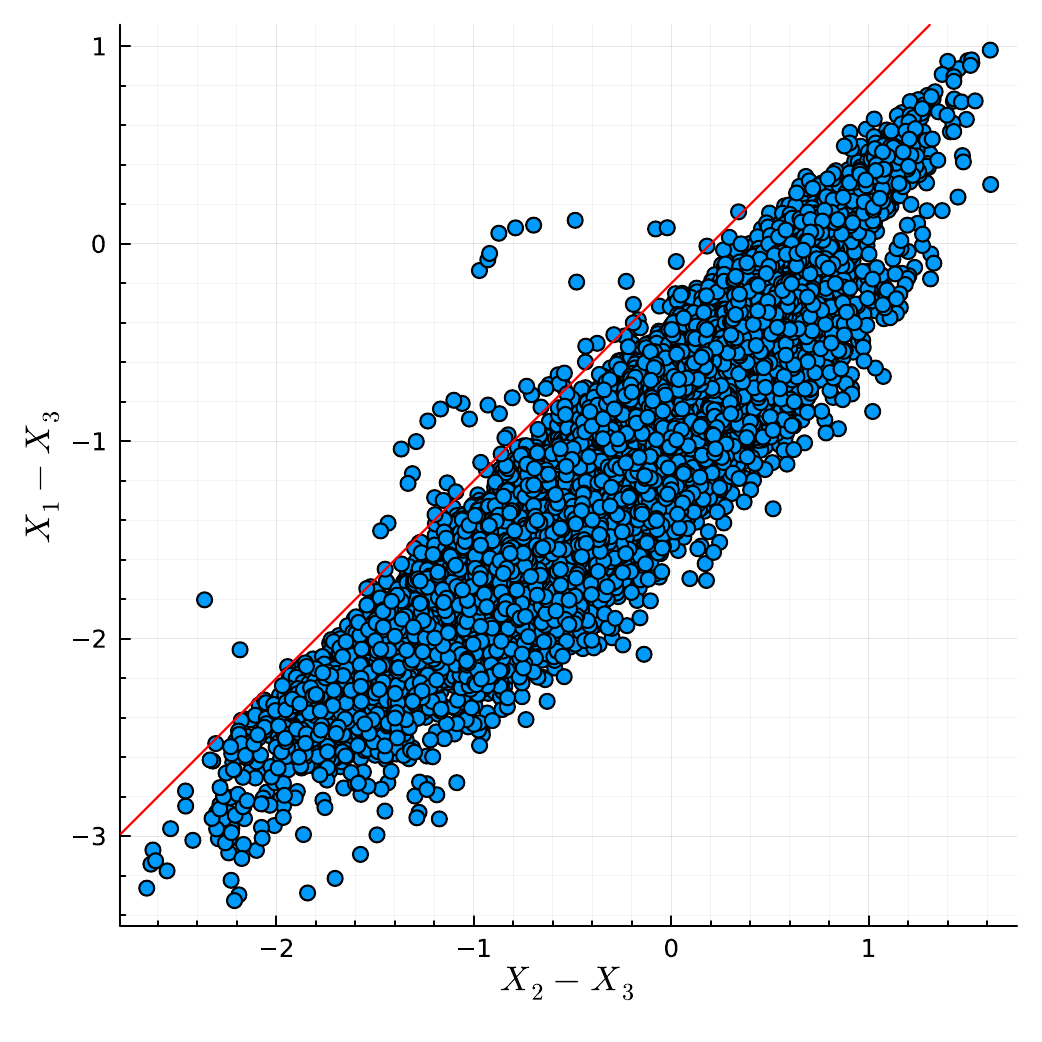}
    \caption{A slice of the upper Danube data plotted in the \(X_{32}\)--\(X_{31}\)-plane. The orange line shows
    the supposed hyperplane corresponding to the edge \(2\to 1\) that is obscured by the noise lying above.}
    \label{fig:danube-slice}
\end{figure}

Inspecting a slice (\Cref{fig:danube-slice}) along the \(X_2-X_3\) and \(X_1 - X_3\) coordinates reveals 
that while there is a visible diagonal line along which the sample concentrates which corresponds 
to the edge \(2\to 1\) there are sample points lying beyond this diagonal line. 
These points interfere with our approach since by calculating \(\widetilde{C}\) we just take the 
maxima without accounting for noise.

\begin{table}[h]
   \centering
   \begin{tabular}{rccccc}
       \toprule
                 & parameters          & nSHD   & FDR    & FPR    & TPR \\
       \midrule
         top-$k$ & $k=203$             & 58.0\% & 64.8\% & 18.9\% & 51.9\% \\
         \midrule
quantile-to-mean & $\overline{r}=0.95$ & 72.3\% & 81.9\% & 52.3\% & 58.4\% \\
       \bottomrule
   \end{tabular}
   \caption{Metrics for the DAGs estimated from the upper Danube network}
   \label{tab:danube}
\end{table}

\section{An open question}
In \Cref{sec:param-recovery}, we studied lower bounds on the size of special samples to completely recover the weights of
an MLBN on a DAG \(\thedag\). We did this in terms of minimal set covers of regular central subdivision of \(\fundamentalpolytope\)
and obtained the minimum sample size \(c(\thedag)\) as the maximum of those set cover numbers.

While we laid the groundwork to study those minimal set covers, there is an immediate questions that serve as next step
in understanding those set covers and the minimum sample size.
\begin{question}
  Is there an expression for the sequence \((c(\kappa_d))_{d\in\NN}\) as recurrence relation or in closed form
  and does it relate to other known sequences?
\end{question}
For example, the entries from \Cref{tab:enumeration} for \(d\leq 7\) suggest that \(c(\kappa_d)\) is the maximum of two numbers,
which is not supported by the evidence for \(d=9\). Also, while closely related to the well-known formula for the 
sum of the first \(d-1\) numbers, \(c(\kappa_d)\) is not just the number of terms in this sum, that is, \(c(\kappa_d)\neq d\).
The difference is that compared to the sum of the first \(d-1\) numbers, a minimum set cover for a subdivision \(\subdivision\)
gives a different expression for \(\frac{(d-1)(d-2)}{2}\) via \[
  \sum_{\sigma\in\subdivision} \#\sigma = \frac{(d-1)(d-2)}{2}
\] where compared to the expression as the sum of the first \(d-1\) numbers several of the simians are combined 
if there is an appropriate \(\sigma\in\subdivision\).

On the other hand, when trying to understand the structures of the minimal set covers, note that any set cover \(\widetilde\subdivision\)
corresponds to a decomposition of \(\thedag\) as a spanning forest where the choice of spanning trees is restricted 
according to \(\subdivision\).
If we assume that \(\subdivision\) is a triangulation, we can even assume that the spanning trees are edge-disjoint.

\appendix
\begin{table}[p]
\begin{subtable}{\linewidth}
   \centering
   \caption{Gaussian i.\,i.\,d.\ innovations and \(G = \kappa_d\)}
   \begin{tabular}{rrlrlrlrl}
       \toprule
       $d$ & \multicolumn{2}{c}{nSHD} & \multicolumn{2}{c}{FDR} & \multicolumn{2}{c}{FPR} & \multicolumn{2}{c}{TPR} \\
       \midrule
         5 &  0.0\% & (0.0\%)         &  0.0\% & (0.0\%)        &  0.0\% & (0.0\%)        & 100.0\% & (0.0\%)       \\
        10 &  0.3\% & (0.2\%)         &  0.3\% & (0.2\%)        &  0.5\% & (0.4\%)        & 99.9\% & (0.3\%)        \\
        30 &  6.5\% & (3.4\%)         &  6.5\% & (3.4\%)        &  7.5\% & (3.9\%)        & 93.4\% & (3.4\%)        \\
        50 & 12.6\% & (3.7\%)         & 12.6\% & (3.7\%)        & 13.7\% & (4.0\%)        & 87.3\% & (3.7\%)        \\
       100 & 17.4\% & (2.8\%)         & 17.4\% & (2.8\%)        & 18.1\% & (2.9\%)        & 82.5\% & (2.8\%)        \\
       \bottomrule
   \end{tabular}
   \label{tab:simulation-study-metrics:gaussian-complete}
\end{subtable}\\\vspace{2em}
\begin{subtable}{\linewidth}
   \centering
   \caption{Gaussian i.\,i.\,d.\ innovations and edge probability \(p = \frac12\)}
   \begin{tabular}{rrlrlrlrl}
       \toprule
       $d$ & \multicolumn{2}{c}{nSHD} & \multicolumn{2}{c}{FDR} & \multicolumn{2}{c}{FPR} & \multicolumn{2}{c}{TPR} \\
       \midrule
         5 &  0.5\% & (0.2\%)         &  0.9\% & (0.3\%)        &  1.1\% & (3.7\%)        & 100.0\% & (0.0\%)       \\
        10 &  0.9\% & (0.1\%)         &  1.9\% & (2.7\%)        &  1.7\% & (2.5\%)        & 99.9\% & (0.1\%)        \\
        30 &  4.7\% & (2.4\%)         &  6.9\% & (2.7\%)        &  6.8\% & (2.7\%)        & 97.6\% & (2.4\%)        \\
        50 & 10.5\% & (4.4\%)         & 12.3\% & (4.3\%)        & 12.3\% & (4.3\%)        & 91.4\% & (4.5\%)        \\
       100 & 18.0\% & (4.6\%)         & 19.0\% & (4.5\%)        & 19.0\% & (4.5\%)        & 83.1\% & (4.7\%)        \\
       \bottomrule
   \end{tabular}
   \label{tab:simulation-study-metrics:gaussian-random-graph}
\end{subtable}\\\vspace{2em}
\begin{subtable}{\linewidth}
   \centering
   \caption{Fr\'echet i.\,i.\,d.\ innovations and \(G = \kappa_d\)}
   \begin{tabular}{rrlrlrlrl}
       \toprule
       $d$ & \multicolumn{2}{c}{nSHD} & \multicolumn{2}{c}{FDR} & \multicolumn{2}{c}{FPR} & \multicolumn{2}{c}{TPR} \\
       \midrule
         5 &  0.0\% & (0.0\%)         &  0.0\% & (0.0\%)        &  0.0\% & (0.0\%)        & 100.0\% & (0.0\%)       \\
        10 &  0.9\% & (1.5\%)         &  0.9\% & (1.5\%)        &  1.5\% & (2.3\%)        & 99.9\% & (0.1\%)        \\
        30 & 14.2\% & (5.0\%)         & 14.2\% & (5.0\%)        & 16.3\% & (5.7\%)        & 85.8\% & (4.9\%)        \\
        50 & 17.8\% & (4.6\%)         & 17.8\% & (4.6\%)        & 19.3\% & (5.0\%)        & 82.2\% & (4.6\%)        \\
       100 & 21.6\% & (3.6\%)         & 21.6\% & (3.6\%)        & 22.5\% & (3.7\%)        & 78.4\% & (3.6\%)        \\
       \bottomrule
   \end{tabular}
   \label{tab:simulation-study-metrics:frechet-complete}
\end{subtable}\\\vspace{2em}
\begin{subtable}{\linewidth}
   \centering
   \caption{Fr\'echet i.\,i.\,d.\ innovations and edge probability \(p = \frac12\)}
   \begin{tabular}{rrlrlrlrl}
       \toprule
       $d$ & \multicolumn{2}{c}{SHD} & \multicolumn{2}{c}{FDR} & \multicolumn{2}{c}{FPR} & \multicolumn{2}{c}{TPR} \\
       \midrule
         5 &  0.0\% & (0.0\%)        &  0.0\% & (0.0\%)        &  0.0\% & (0.0\%)        & 100.0\% & (0.0\%)       \\
        10 &  2.3\% & (1.8\%)        &  4.4\% & (3.3\%)        &  4.4\% & (3.3\%)        &  99.9\% & (0.5\%)       \\
        30 & 14.8\% & (4.9\%)        & 17.9\% & (4.8\%)        & 18.1\% & (4.9\%)        &  88.5\% & (5.1\%)       \\
        50 & 20.5\% & (6.1\%)        & 22.5\% & (6.0\%)        & 22.7\% & (6.0\%)        &  81.7\% & (6.3\%)       \\
       100 & 23.4\% & (5.2\%)        & 24.5\% & (5.1\%)        & 24.7\% & (5.2\%)        &  77.7\% & (5.3\%)       \\
       \bottomrule
   \end{tabular}
   \label{tab:simulation-study-metrics:frechet-random-graph}
\end{subtable}
\caption{Simulation study for fixed topological ordering and top-$k$ gap scoring where \(k=30\)}
\label{tab:simulation-study-metrics:fixed-ordering}
\end{table}

\begin{table}[p]
\begin{subtable}{\linewidth}
   \centering
   \caption{Gaussian i.\,i.\,d.\ innovations and \(G = \kappa_d\)}
   \begin{tabular}{rrlrlrlrl}
       \toprule
       $d$ & \multicolumn{2}{c}{SHD} & \multicolumn{2}{c}{FDR} & \multicolumn{2}{c}{FPR} & \multicolumn{2}{c}{TPR} \\
       \midrule
         5 &  0.0\% & (0.0\%)        &  0.0\% & (0.0\%)        &  0.0\% & (0.0\%)        & 100.0\% & (0.0\%)       \\
        10 &  0.0\% & (0.3\%)        &  0.0\% & (0.3\%)        &  0.1\% & (0.4\%)        &  99.9\% & (0.2\%)        \\
        30 &  9.9\% & (4.6\%)        &  9.9\% & (4.6\%)        & 11.4\% & (5.3\%)        &  90.1\% & (4.6\%)        \\
        50 & 21.2\% & (4.9\%)        & 21.2\% & (4.9\%)        & 23.0\% & (5.3\%)        &  78.8\% & (4.9\%)        \\
       100 & 33.6\% & (3.7\%)        & 33.7\% & (3.7\%)        & 35.1\% & (3.9\%)        &  66.3\% & (3.7\%)        \\
       \bottomrule
   \end{tabular}
   \label{tab:simulation-study-metrics:gaussian-complete}
\end{subtable}\\\vspace{2em}
\begin{subtable}{\linewidth}
   \centering
   \caption{Gaussian i.\,i.\,d.\ innovations and edge probability \(p = \frac12\)}
   \begin{tabular}{rrlrlrlrl}
       \toprule
       $d$ & \multicolumn{2}{c}{SHD} & \multicolumn{2}{c}{FDR} & \multicolumn{2}{c}{FPR} & \multicolumn{2}{c}{TPR} \\
       \midrule
         5 &  0.0\% & (0.0\%)        &  0.0\% & (0.0\%)        &  0.0\% & (0.0\%)        & 100.0\% & (0.0\%)       \\
        10 &  0.9\% & (1.4\%)        &  1.7\% & (2.6\%)        &  1.6\% & (2.3\%)        &  99.9\% & (0.1\%)        \\
        30 &  5.0\% & (2.6\%)        &  7.2\% & (2.9\%)        &  7.1\% & (2.9\%)        &  97.3\% & (2.7\%)        \\
        50 & 13.2\% & (5.5\%)        & 14.9\% & (5.3\%)        & 14.9\% & (5.4\%)        &  88.7\% & (5.7\%)        \\
       100 & 27.6\% & (5.6\%)        & 28.5\% & (5.5\%)        & 28.5\% & (5.6\%)        &  73.4\% & (5.7\%)        \\
       \bottomrule
   \end{tabular}
   \label{tab:simulation-study-metrics:gaussian-random-graph}
\end{subtable}\\\vspace{2em}
\begin{subtable}{\linewidth}
   \centering
   \caption{Fr\'echet i.\,i.\,d.\ innovations and \(G = \kappa_d\)}
   \begin{tabular}{rrlrlrlrl}
       \toprule
       $d$ & \multicolumn{2}{c}{SHD} & \multicolumn{2}{c}{FDR} & \multicolumn{2}{c}{FPR} & \multicolumn{2}{c}{TPR} \\
       \midrule
         5 &  0.0\% & (0.0\%)        &  0.0\% & (0.0\%)        &  0.0\% & (0.0\%)        & 100.0\% & (0.0\%)       \\
        10 &  1.0\% & (1.5\%)        &  1.0\% & (1.5\%)        &  1.6\% & (2.4\%)        &  99.0\% & (1.5\%)       \\
        30 & 21.2\% & (5.5\%)        & 21.2\% & (5.5\%)        & 24.9\% & (6.3\%)        &  78.3\% & (5.5\%)       \\
        50 & 31.5\% & (5.2\%)        & 31.5\% & (5.2\%)        & 34.2\% & (5.6\%)        &  68.5\% & (5.2\%)       \\
       100 & 40.1\% & (3.8\%)        & 40.1\% & (3.8\%)        & 41.7\% & (3.9\%)        &  59.9\% & (3.8\%)       \\
       \bottomrule
   \end{tabular}
   \label{tab:simulation-study-metrics:frechet-complete}
\end{subtable}\\\vspace{2em}
\begin{subtable}{\linewidth}
   \centering
   \caption{Fr\'echet i.\,i.\,d.\ innovations and edge probability \(p = \frac12\)}
   \begin{tabular}{rrlrlrlrl}
       \toprule
       $d$ & \multicolumn{2}{c}{SHD} & \multicolumn{2}{c}{FDR} & \multicolumn{2}{c}{FPR} & \multicolumn{2}{c}{TPR} \\
       \midrule
         5 &  0.0\% & (0.0\%)        &  0.0\% & (0.0\%)        &  0.0\% & (0.0\%)        & 100.0\% & (0.0\%)       \\
        10 &  2.3\% & (1.7\%)        &  4.4\% & (3.3\%)        &  4.4\% & (3.3\%)        &  99.9\% & (0.5\%)       \\
        30 & 17.2\% & (5.5\%)        & 20.3\% & (5.3\%)        & 20.5\% & (5.5\%)        &  86.0\% & (5.8\%)       \\
        50 & 27.6\% & (6.6\%)        & 29.5\% & (6.4\%)        & 29.8\% & (6.5\%)        &  74.4\% & (6.8\%)       \\
       100 & 37.9\% & (5.5\%)        & 38.8\% & (5.4\%)        & 39.1\% & (5.5\%)        &  63.0\% & (5.5\%)        \\
       \bottomrule
   \end{tabular}
   \label{tab:simulation-study-metrics:frechet-random-graph}
\end{subtable}
\caption{Simulation study for random topological ordering and top-$k$ gap scoring where \(k=30\)}
\label{tab:simulation-study-metrics:random-ordering}
\end{table}

\begin{table}[p]
\begin{subtable}{\linewidth}
   \centering
   \caption{Gaussian i.\,i.\,d.\ innovations and \(G = \kappa_d\)}
   \begin{tabular}{rrlrlrlrl}
       \toprule
       $d$ & \multicolumn{2}{c}{nSHD} & \multicolumn{2}{c}{FDR} & \multicolumn{2}{c}{FPR} & \multicolumn{2}{c}{TPR} \\
       \midrule
         5 &  0.9\% & (2.4\%)         &  0.9\% & (2.4\%)        &  2.8\% & (7.1\%)        & 99.1\% & (2.4\%)        \\
        10 &  8.9\% & (4.9\%)         &  8.9\% & (4.9\%)        & 14.0\% & (7.5\%)        & 91.1\% & (4.8\%)        \\
        30 & 34.8\% & (7.2\%)         & 34.8\% & (7.2\%)        & 40.0\% & (8.3\%)        & 65.2\% & (7.2\%)        \\
        50 & 40.7\% & (5.9\%)         & 40.7\% & (5.9\%)        & 44.2\% & (6.4\%)        & 59.3\% & (5.9\%)        \\
       100 & 44.4\% & (3.4\%)         & 44.4\% & (3.4\%)        & 46.2\% & (3.6\%)        & 55.6\% & (3.4\%)        \\
       \bottomrule
   \end{tabular}
   \label{tab:simulation-study-metrics:gaussian-complete}
\end{subtable}\\\vspace{2em}
\begin{subtable}{\linewidth}
   \centering
   \caption{Gaussian i.\,i.\,d.\ innovations and edge probability \(p = \frac12\)}
   \begin{tabular}{rrlrlrlrl}
       \toprule
       $d$ & \multicolumn{2}{c}{nSHD} & \multicolumn{2}{c}{FDR} & \multicolumn{2}{c}{FPR} & \multicolumn{2}{c}{TPR} \\
       \midrule
         5 & 15.5\% & (9.7\%)         & 25.6\% & (14.4\%)       & 39.7\% & (16.0\%)       & 99.9\% & (0.9\%)       \\
        10 & 14.3\% & (5.4\%)         & 23.1\% & (8.3\%)        & 26.6\% & (7.0\%)        & 97.5\% & (3.2\%)        \\
        30 & 24.0\% & (7.8\%)         & 27.6\% & (7.3\%)        & 28.7\% & (7.7\%)        & 79.9\% & (8.5\%)        \\
        50 & 33.6\% & (8.3\%)         & 35.6\% & (8.0\%)        & 36.3\% & (8.2\%)        & 68.5\% & (8.6\%)        \\
       100 & 41.3\% & (6.4\%)         & 42.2\% & (6.3\%)        & 42.6\% & (6.4\%)        & 59.6\% & (6.5\%)        \\
       \bottomrule
   \end{tabular}
   \label{tab:simulation-study-metrics:gaussian-random-graph}
\end{subtable}\\\vspace{2em}
\begin{subtable}{\linewidth}
   \centering
   \caption{Fr\'echet i.\,i.\,d.\ innovations and \(G = \kappa_d\)}
   \begin{tabular}{rrlrlrlrl}
       \toprule
       $d$ & \multicolumn{2}{c}{nSHD} & \multicolumn{2}{c}{FDR} & \multicolumn{2}{c}{FPR} & \multicolumn{2}{c}{TPR} \\
       \midrule
         5 & 11.1\% & (12.6\%)        & 10.8\% & (12.6\%)       & 32.3\% & (37.8\%)       & 88.6\% & (12.9\%)       \\
        10 & 22.3\% & (10.2\%)        & 22.1\% & (10.2\%)       & 34.5\% & (16.0\%)       & 77.6\% & (10.3\%)       \\
        30 & 39.7\% & (7.4\%)         & 39.7\% & (7.4\%)        & 45.5\% & (8.5\%)        & 60.2\% & (7.4\%)        \\
        50 & 42.6\% & (5.4\%)         & 42.5\% & (5.4\%)        & 46.1\% & (5.8\%)        & 57.4\% & (5.4\%)        \\
       100 & 44.4\% & (3.0\%)         & 44.4\% & (3.0\%)        & 46.2\% & (3.1\%)        & 55.6\% & (3.0\%)        \\
       \bottomrule
   \end{tabular}
   \label{tab:simulation-study-metrics:frechet-complete}
\end{subtable}\\\vspace{2em}
\begin{subtable}{\linewidth}
   \centering
   \caption{Fr\'echet i.\,i.\,d.\ innovations and edge probability \(p = \frac12\)}
   \begin{tabular}{rrlrlrlrl}
       \toprule
       $d$ & \multicolumn{2}{c}{SHD} & \multicolumn{2}{c}{FDR} & \multicolumn{2}{c}{FPR} & \multicolumn{2}{c}{TPR} \\
       \midrule
         5 & 24.4\% & (14.2\%)       & 32.9\% & (16.4\%)       & 53.5\% & (23.6\%)       & 88.6\% & (13.7\%)       \\
        10 & 24.9\% & (10.6\%)       & 32.3\% & (11.2\%)       & 37.5\% & (11.5\%)       & 85.0\% & (11.4\%)       \\
        30 & 36.0\% & (9.4\%)        & 38.9\% & (8.9\%)        & 40.3\% & (9.3\%)        & 67.2\% & (10.0\%)       \\
        50 & 40.5\% & (8.7\%)        & 42.2\% & (8.4\%)        & 43.0\% & (8.6\%)        & 61.3\% & (9.0\%)        \\
       100 & 43.9\% & (5.8\%)        & 44.8\% & (5.7\%)        & 45.2\% & (5.8\%)        & 56.9\% & (5.9\%)        \\
       \bottomrule
   \end{tabular}
   \label{tab:simulation-study-metrics:frechet-random-graph}
\end{subtable}
\caption{Simulation study for fixed topological ordering and quantile-to-mean scoring with upper quantile level 
\(\overline{r}=0.95\)}
\label{tab:simulation-study-metrics:fixed-ordering:quantile-to-mean}
\end{table}

\begin{table}[p]
\begin{subtable}{\linewidth}
   \centering
   \caption{Gaussian i.\,i.\,d.\ innovations and \(G = \kappa_d\)}
   \begin{tabular}{rrlrlrlrl}
       \toprule
       $d$ & \multicolumn{2}{c}{nSHD} & \multicolumn{2}{c}{FDR} & \multicolumn{2}{c}{FPR} & \multicolumn{2}{c}{TPR} \\
       \midrule
         5 &  0.9\% & (2.4\%)         &  0.9\% & (2.4\%)        &  2.8\% & (7.1\%)        & 99.1\% & (2.4\%)        \\
        10 &  9.1\% & (4.9\%)         &  9.1\% & (4.9\%)        & 14.3\% & (7.7\%)        & 91.1\% & (4.9\%)        \\
        30 & 35.2\% & (7.5\%)         & 35.2\% & (7.5\%)        & 40.4\% & (8.7\%)        & 64.8\% & (7.5\%)        \\
        50 & 41.8\% & (5.9\%)         & 41.8\% & (5.9\%)        & 45.3\% & (6.4\%)        & 58.2\% & (5.9\%)        \\
       100 & 45.9\% & (3.5\%)         & 45.9\% & (3.5\%)        & 47.8\% & (3.7\%)        & 54.0\% & (3.5\%)        \\
       \bottomrule
   \end{tabular}
   \label{tab:simulation-study-metrics:gaussian-complete}
\end{subtable}\\\vspace{2em}
\begin{subtable}{\linewidth}
   \centering
   \caption{Gaussian i.\,i.\,d.\ innovations and edge probability \(p = \frac12\)}
   \begin{tabular}{rrlrlrlrl}
       \toprule
       $d$ & \multicolumn{2}{c}{nSHD} & \multicolumn{2}{c}{FDR} & \multicolumn{2}{c}{FPR} & \multicolumn{2}{c}{TPR} \\
       \midrule
         5 & 15.5\% & (9.8\%)         & 25.6\% & (14.5\%)       & 39.7\% & (16.1\%)       & 99.9\% & (0.9\%)       \\
        10 & 14.2\% & (5.4\%)         & 22.9\% & (8.3\%)        & 26.4\% & (7.0\%)        & 97.5\% & (3.1\%)        \\
        30 & 24.1\% & (7.6\%)         & 27.7\% & (7.1\%)        & 28.8\% & (7.5\%)        & 79.8\% & (8.3\%)        \\
        50 & 34.0\% & (8.4\%)         & 36.0\% & (8.1\%)        & 36.7\% & (8.4\%)        & 68.1\% & (8.8\%)        \\
       100 & 42.0\% & (6.6\%)         & 42.9\% & (6.5\%)        & 43.3\% & (6.5\%)        & 58.9\% & (6.7\%)        \\
       \bottomrule
   \end{tabular}
   \label{tab:simulation-study-metrics:gaussian-random-graph}
\end{subtable}\\\vspace{2em}
\begin{subtable}{\linewidth}
   \centering
   \caption{Fr\'echet i.\,i.\,d.\ innovations and \(G = \kappa_d\)}
   \begin{tabular}{rrlrlrlrl}
       \toprule
       $d$ & \multicolumn{2}{c}{nSHD} & \multicolumn{2}{c}{FDR} & \multicolumn{2}{c}{FPR} & \multicolumn{2}{c}{TPR} \\
       \midrule
         5 & 11.0\% & (12.5\%)        & 10.7\% & (12.5\%)       & 31.9\% & (37.4\%)       & 88.8\% & (12.7\%)       \\
        10 & 22.3\% & (10.0\%)        & 22.1\% & (10.0\%)       & 34.7\% & (15.7\%)       & 77.6\% & (10.1\%)       \\
        30 & 41.3\% & (7.6\%)         & 41.3\% & (7.6\%)        & 47.3\% & (8.7\%)        & 58.6\% & (7.6\%)        \\
        50 & 44.7\% & (5.4\%)         & 44.7\% & (5.4\%)        & 48.4\% & (5.9\%)        & 55.3\% & (5.4\%)        \\
       100 & 47.3\% & (3.1\%)         & 47.3\% & (3.1\%)        & 49.2\% & (3.2\%)        & 52.6\% & (3.1\%)        \\
       \bottomrule
   \end{tabular}
   \label{tab:simulation-study-metrics:frechet-complete}
\end{subtable}\\\vspace{2em}
\begin{subtable}{\linewidth}
   \centering
   \caption{Fr\'echet i.\,i.\,d.\ innovations and edge probability \(p = \frac12\)}
   \begin{tabular}{rrlrlrlrl}
       \toprule
       $d$ & \multicolumn{2}{c}{SHD} & \multicolumn{2}{c}{FDR} & \multicolumn{2}{c}{FPR} & \multicolumn{2}{c}{TPR} \\
       \midrule
         5 & 25.1\% & (14.3\%)       & 33.8\% & (16.4\%)       & 54.4\% & (23.2\%)       & 88.6\% & (13.8\%)       \\
        10 & 25.0\% & (10.5\%)       & 32.5\% & (11.0\%)       & 37.8\% & (11.4\%)       & 84.9\% & (11.3\%)       \\
        30 & 36.6\% & (9.5\%)        & 39.5\% & (9.0\%)        & 40.9\% & (9.4\%)        & 66.6\% & (10.2\%)       \\
        50 & 41.8\% & (8.7\%)        & 43.4\% & (8.5\%)        & 44.2\% & (8.6\%)        & 60.0\% & (9.1\%)        \\
       100 & 45.5\% & (6.0\%)        & 46.3\% & (5.9\%)        & 46.7\% & (6.0\%)        & 55.3\% & (6.1\%)        \\
       \bottomrule
   \end{tabular}
   \label{tab:simulation-study-metrics:frechet-random-graph}
\end{subtable}
\caption{Simulation study for random topological ordering and quantile-to-mean scoring with upper quantile level 
\(\overline{r}=0.95\)}
\label{tab:simulation-study-metrics:random-ordering:quantile-to-mean}
\end{table}

\section*{Acknowledgements}
The author is funded by the Deutsche Forschungsgemeinschaft (DFG, German Research Foundation) under Germany´s Excellence Strategy 
– The Berlin Mathematics Research Center MATH+ (EXC-2046/1, project ID: 390685689), Project AA3-16.
\singlespace
\printbibliography

\end{document}